\documentclass[reprint,amssymb,amsmath,twocolumn,aps,prl,showpacs,groupedaddress,footinbib,longbibliography,superscriptaddress]{revtex4-1}

\usepackage[T1]{fontenc}
\usepackage[utf8]{inputenc}
\usepackage{textcomp}

\usepackage{graphicx}
\usepackage{color}
\usepackage[colorlinks,bookmarks=false,citecolor=darkblue,linkcolor=red,urlcolor=blue]{hyperref}

\definecolor{darkred}{rgb}{0.7,0.0,0.0}

\definecolor{darkblue}{rgb}{0,0.02,0.45}

\definecolor{darkgreen}{rgb}{0.02,0.45,0.0}

\definecolor{violet}{rgb}{0.8,0.2,0.6}

\usepackage{float}
\usepackage{color}
\usepackage{bm}
\usepackage{amsmath}
\usepackage{amssymb}
\usepackage{graphicx}
\usepackage{epstopdf}

\makeatletter

\usepackage{latexsym}
\usepackage[english]{babel}

\begin{document}

\title{Thermodynamics of the kagome-lattice Heisenberg antiferromagnet with arbitrary spin $S$.}

\author{P. M\"uller}
\affiliation{Institut f\"ur Theoretische Physik,
Otto-von-Guericke-Universit\"at Magdeburg,
D-39016 Magdeburg, Germany}

\author{A. Zander}
\affiliation{Institut f\"ur Theoretische Physik,
Otto-von-Guericke-Universit\"at Magdeburg,
D-39016 Magdeburg, Germany}

\author{J. Richter}
\affiliation{Institut f\"ur Theoretische Physik,
Otto-von-Guericke-Universit\"at Magdeburg,
D-39016 Magdeburg, Germany}
\affiliation{Max-Planck-Institut f\"{u}r Physik Komplexer Systeme,
        N\"{o}thnitzer Stra{\ss}e 38, 01187 Dresden, Germany}

\date{\today}
\begin{abstract}
We use a second-order rotational invariant Green's function method (RGM) and the
high-temperature expansion (HTE) to calculate the thermodynamic properties,
of
the kagome-lattice spin-$S$ Heisenberg antiferromagnet with nearest-neighbor exchange
$J$.
While the HTE yields accurate results down to temperatures of about
$T/S(S+1) \sim J$, the RGM provides data for arbitrary $T \ge 0$.
For the ground state we use the RGM data to  analyze the $S$-dependence of
the excitation spectrum, the
excitation velocity, the uniform susceptibility, the spin-spin correlation functions, the correlation
length, and the structure factor. We found that the so-called
$\sqrt{3}\times\sqrt{3}$ ordering is more pronounced than the
$q=0$ ordering for all values of $S$. In the extreme quantum case $S=1/2$ the
zero-temperature  correlation
length is only of the order of the nearest-neighbor separation.        
Then we study the temperature dependence of several physical quantities for spin
quantum numbers $S=1/2,1,\dots,7/2$. 
As increasing $S$
the typical maximum in the specific heat and in the uniform susceptibility are
shifted towards lower values of $T/S(S+1)$ 
and the height of the maximum is growing.
The structure factor ${\cal S}(\mathbf{q})$ exhibits 
two maxima at magnetic wave vectors $\mathbf{q}={\mathbf{Q}_i}, i=0,1,$
corresponding to the $q=0$ and $\sqrt{3}\times\sqrt{3}$ state.
We find that the  $\sqrt{3}\times \sqrt{3}$ short-range order  is more
pronounced 
than the $q=0$ short-range order   
for all temperatures $T \ge 0$.
For the spin-spin correlation functions, the correlation lengths and the structure factors,
we find a finite low-temperature region $0 \le T < T^*\approx a/S(S+1)$, $a
\approx 0.2$,  
where these quantities are almost independent of $T$.
\end{abstract}
\maketitle

\section{Introduction\protect \\
}
\label{sec:intro}

One of the most prominent and at the same time challenging spin
models with a frustration induced highly
degenerated classical ground state (GS) manifold is the kagome Heisenberg 
antiferromagnet (KHAF)
\cite{Elser1990,Chalker1992,Harris1992,Singh1992,Reimers1993,elstner1993,elstner1994,Henley:1995,Nakamura1995,tomczak1996thermodynamical,Waldtmann1998,Yu2000,Lhuillier_thermo_PRL2000,Huber2001,Bernhard2002,Schmalfus2004,Bernu2005,Singh2007,Li2007,Rigol2007,Zhitomirsky2008,DMRG_PRL08,Laeuchli2009,Evenbly2010,Goetze2011,Nakano2011,Iqbal2011,Yan2011,Laeuchli2011,Depenbrock2012,Rousochatzakis2013,
Iqbal2013,Rousochatzakis2014,Xie2014,Lohmann2014,Munehisha2014,Kolley2015,Goetze2015,Changlani2015,Liu2015,Picot2015,Nishimoto2015,Liu2016,Shimokawa2016,Oitmaa2016,Goetze2016,Laeuchli2016,Pollmann2017,Xie2017,Singh2017,Xi-Chen2017}.
This degeneracy is lifted by fluctuations ({\it order from disorder}
mechanism)
\cite{villain,shender2,Henley:1995}.
Particular attention has been paid to the extreme quantum spin-half case  
[\onlinecite{Waldtmann1998,Yu2000,Lhuillier_thermo_PRL2000,Bernhard2002,Schmalfus2004,Singh2007,Li2007,Zhitomirsky2008,DMRG_PRL08,Laeuchli2009,Evenbly2010,Goetze2011,Nakano2011,Iqbal2011,Yan2011,Laeuchli2011,
Depenbrock2012,Rousochatzakis2013,Iqbal2013,Rousochatzakis2014,Xie2014,Lohmann2014,Kolley2015,Goetze2015,Goetze2016,Laeuchli2016,Pollmann2017,Xie2017,Singh2017,Xi-Chen2017}].
Although, there is consensus on the absence of magnetic long-range order 
(LRO) the nature of the spin-liquid GS is still under debate.
Meanwhile also for spin quantum number $S=1$ there is evidence that the KHAF
does not exhibit magnetic LRO
\cite{Goetze2011,Goetze2015,Changlani2015,Liu2015,Picot2015,Nishimoto2015}.  
Recently it has been argued that there is a route to magnetic
GS
LRO in the KHAF
as increasing the spin quantum number to $S\ge3/2$, see
\cite{Goetze2011,Goetze2015,Oitmaa2016,Liu2016}.

Except the theoretical work there is also a large activity on the experimental
side.
Among the $S=1/2$ kagome compounds, Herbertsmithite ZnCu$_3$(OH)$_6$Cl$_2$ is
a promising candidate for a spin liquid, see
\cite{herbertsmithite2007,herbertsmithite2007a,Hiroi2009,herbertsmithite2009,herbertsmithite2010,herbertsmithite2012}.
Examples
for kagome magnets with higher spin $S$ are deuteronium jarosite
(D$_3$O)Fe$_3$(SO$_4$)$_2$(OD)$_6$ with spin $S=5/2$, see \cite{Fak_2007}, and
 the recently studied Cr-Jarosite KCr$_3$(OH)$_6$(SO$_4$)$_2$ with spin $S=3/2$, see
\cite{Okubo2017}.
\begin{figure}
\hspace*{-0.5cm}
\centering 
\includegraphics[scale=1]{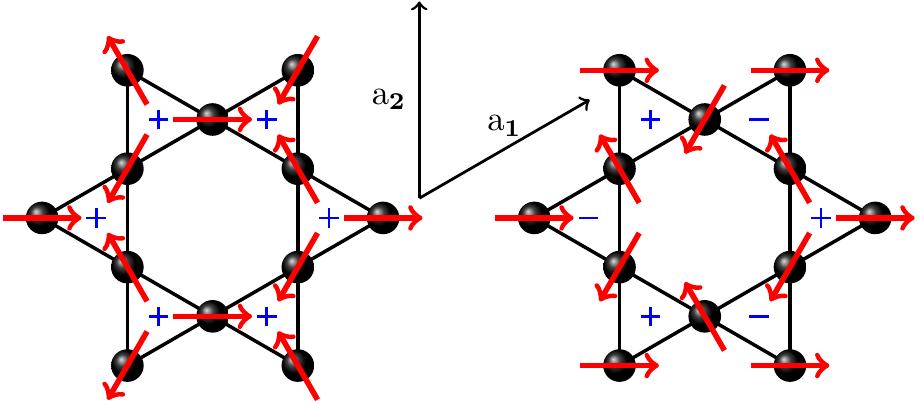} 
\protect\caption{
Illustration of the two most relevant classical states. Left: $q=0$ state
with magnetic wave vector $\mathbf{Q}_0=(2\pi/\sqrt{3},0)$. Right: $\sqrt{3}\times\sqrt{3}$ state with magnetic wave
vector $\mathbf{Q}_1=(0,4\pi/3)$. 
"$+$" and "$-$" symbols denote plaquettes of different vector spin chirality.
The arrows indicate the basis vectors $\mathbf{a}_1$ and $\mathbf{a}_2$. 
}
\label{Fig1} 
\end{figure}
Due to the
{\it order from disorder} mechanism two different coplanar states may be
selected by fluctuations:
(i) The so called $q=0$ state with a corresponding magnetic wave vector $\mathbf{Q}_0=(2\pi/\sqrt{3},0)$
(Fig.~\ref{Fig1}, left), which has a  magnetic unit cell that is identical to
the geometrical one.
(ii) The so called $\sqrt{3}\times\sqrt{3}$ state (Fig.~\ref{Fig1}, right) with a corresponding
magnetic wave vector $\mathbf{Q}_1=(0,4\pi/3)$ 
which has a three times larger unit cell, cf. e.g., 
\cite{Zhitomirsky2008}.
Moreover, both states are characterized by 
different vector chirality patterns, see Fig. \ref{Fig1}.
The selection of one of these states is a subtle issue and depends on spin
quantum number, anisotropy etc., see, e.g.,
\cite{Sachdev_1992,Chubukov:92,Henley:1995,Goetze2011,Zhito_PRL_XXZ,Chernyshev2015,Goetze2015,Goetze2016}.       
While for the widely studied GS properties a plethora of many-body methods
are available, the tool box for the calculation of finite-temperature
properties of highly
frustrated quantum magnets is sparse.
Here we use two universal approaches suitable 
to calculate thermodynamic quantities of Heisenberg quantum spin systems of arbitrary lattice geometry,
namely 
the Green-function technique \cite{Gasser2001,Nolting2009,Froebrich2006} 
and 
the high-temperature expansion
\cite{elstner1993,elstner1994,2DJ1J2,singh2012,Kapellasite,Oitmaa2006,Bernu2001,Bernu2005,Bernu2015,Lohmann2011,Lohmann2014,Richter2015,Schmidt2017,Singh2017}.

We study the kagome lattice with antiferromagnetic ($J>0$) nearest-neighbor interaction
\begin{eqnarray}\label{model}
\hat{H}  =  J \hspace{-2mm} \sum_{\langle
m\alpha,n\beta\rangle}\bm{\hat{S}}_{m\alpha}\bm{\hat{S}}_{n\beta}\; , \;
\bm{\hat{S}}_{m\alpha}^2 =S(S+1), \label{eq_ham}
\end{eqnarray}
where the Greek indices ($\alpha, \beta=1,2,3$) run over the spins in a
geometrical unit cell (that contains three sites) and 
the latin indices $n$ and $m$ label the unit cells given by the basis
vectors $\mathbf{a}_1=(0,2)$ and $\mathbf{a}_2=(\sqrt{3},1)$.

The paper is organized  as follows: In Sec. \ref{methods} we briefly
illustrate the applied methods. In Sec. \ref{sec:GS} we describe  the properties of the model at zero temperature,
followed by the discussion of finite-temperature properties  of the model in Sec. \ref{finite_T}.
In Sec.~\ref{sec:sum} we summarize our findings.

\section{Methods}
\label{methods}

\subsection{Rotation-invariant Green's function method (RGM) \label{RGM}}
A rotation-invariant formalism of the Green's function method was first introduced by Kondo and Yamaji \cite{Kondo1972}
 to describe short-range order (SRO) of the one-dimensional $S=1/2$ Heisenberg ferromagnet at $T>0$.
They decoupled the hierarchy of equation of motions in second order,
i.e., one step beyond 
the usual random-phase approximation (RPA)
\cite{Tyablikov1967,Gasser2001,Nolting2009}
and introduced  rotational invariance by setting $\langle \hat{S}^z_{i}\rangle=0$ in the equations of
motions. 
Within this rotation-invariant scheme
possible magnetic LRO is described by the long-range part in
the two-point spin correlators.
Furthermore, the 
approximation made by the decoupling of higher-order correlators is improved   
by introducing so-called vertex parameters, see below.
In the following decades the rotation-invariant Green's function method (RGM) was further elaborated
to include arbitrary  spin $S$, antiferromagnetic spin systems including
frustrated ones and also
more complex spin-lattices with non-primitive unit cells
\cite{Rhodes1973,Shimahara1991,Suzuki1994,Barabanov94,Ihle1997,Ihle1999,Yu2000,Ihle2001,Bernhard2002,Schmalfus2004,Schmalfus2006,
Schmalfus2005,Junger2005,Juh_sz_Junger_2009,Haertel2008,Haertel2010,Haertel2011,Haertel2011a,Haertel2013,Antsygina2008,Mikheyenkov2013,Vladimirov2014,Mueller2015,Mikheyenkov2016,
Vladimirov2017,2017Mullera,2017Muller}. 
At the present time the RGM is a well established method 
and has been successfully used in numerous recent publications 
on the theory of frustrated spin systems 
\cite{Yu2000,Ihle2001,Bernhard2002,Schmalfus2004,Schmalfus2006,Schmalfus2005,Haertel2008,Haertel2010,Haertel2011,Haertel2011a,Haertel2013,Mueller2015,Junger2005,Juh_sz_Junger_2009,Mikheyenkov2013,Mikheyenkov2016,2017Mullera,2017Muller}.

The early papers using the RGM \cite{Yu2000,Bernhard2002,Schmalfus2004} to study the KHAF
were restricted to the  spin-$1/2$ case and used a simple minimal version of
the RGM, see below. 
In the present paper we extend the RGM approach to arbitrary values of the spin
quantum number $S\ge1$ and improve the previous RGM studies going beyond the  minimal version 
by introducing one more vertex parameter. 
Moreover, we provide  a more comprehensive analysis of the thermodynamic
quantities  by considering, e.g. the temperature dependence of  the
structure factor and correlation lengths.

The basic quantity that has to be determined within the RGM is the (retarded) Green's function
$\langle\langle\hat{S}_{\mathbf{q}\alpha}^{+};\hat{S}_{\mathbf{\mathbf{q}\beta}}^{-}\rangle\rangle_{\omega}$,
which is related to the dynamic wavelength-dependent susceptibility 
$\langle\langle\hat{S}_{\mathbf{q}\alpha}^{+};\hat{S}_{\mathbf{\mathbf{q}\beta}}^{-}\rangle\rangle_{\omega}=-\chi^{+-}_{\alpha\beta\mathbf{q}}(\omega)$.
To determine 
$\langle\langle\hat{S}_{\mathbf{q}\alpha}^{+};\hat{S}_{\mathbf{\mathbf{q}\beta}}^{-}\rangle\rangle_{\omega}$
we use the equation of motion (EoM) up to second order,
\begin{eqnarray}
 \omega\langle\langle\hat{S}_{\mathbf{q}\alpha}^{+};\hat{S}_{\mathbf{\mathbf{q}\beta}}^{-}\rangle\rangle_{\omega}&=&\langle[\hat{S}_{\mathbf{q}\alpha}^{+},\hat{S}_{\mathbf{\mathbf{q}\beta}}^{-}]_{-}\rangle+\langle\langle\textrm{i}\dot{\hat{S}}_{\mathbf{q}\alpha}^{+};\hat{S}_{\mathbf{\mathbf{q}\beta}}^{-}\rangle\rangle_{\omega},\nonumber \\
\omega\langle\langle\textrm{i}\dot{\hat{S}}_{\mathbf{q}\alpha}^{+};\hat{S}_{\mathbf{\mathbf{q}\beta}}^{-}\rangle\rangle_{\omega}&=&\langle[\textrm{i}\dot{\hat{S}}_{\mathbf{q}\alpha}^{+},\hat{S}_{\mathbf{\mathbf{q}\beta}}^{-}]_{-}\rangle-\langle\langle\ddot{\hat{S}}_{\mathbf{q}\alpha}^{+};\hat{S}_{\mathbf{\mathbf{q}\beta}}^{-}\rangle\rangle_{\omega}.
 \label{EoM}
\end{eqnarray}
Naturally, for an interacting many-body problem  more complicated (i.e., higher-order) Green's functions
appear in the EoM.
It is in order to mention here that the RPA, that can be obtained
by applying the EoM only once (first line in Eq.~(\ref{EoM})), has
the disadvantage that only phases with magnetic LRO can be described
properly, since the Green's function is
proportional to magnetic order parameters \cite{Tyablikov1967,Gasser2001,Nolting2009}. 
In contrast, SRO can be adequately described by the RGM due to including the next order in the EoM, see the second line in
Eq.~(\ref{EoM}).
The operator
$\ddot{\hat{S}}_{\mathbf{q}\alpha}^{+}$ appearing in second-order contains several combinations of
three-spin operators.
These products of three-spin operators are simplified by the decoupling scheme along the lines of, e.g., \cite{Suzuki1994,Juh_sz_Junger_2009,Haertel2011,2017Muller}
which can be sketched as follows:

\begin{eqnarray}
\label{decoup_RGM}
\hat{S}_A^-\hat{S}_B^+\hat{S}_C^+ & \rightarrow & \alpha^{}_{AB} c^{+-}_{AB}\hat{S}_C^+ + \alpha^{}_{AC}c^{+-}_{AC}S_B^+,\\ \nonumber
\hat{S}_A^z\hat{S}_B^z\hat{S}_C^+ & \rightarrow & \frac 1 2 \alpha^{}_{AB}c^{+-}_{AB}\hat{S}_C^+,\\ \nonumber 
\hat{S}_A^z\hat{S}_A^z\hat{S}_B^+ & \rightarrow & c^{zz}_{AA}\hat{S}_B^+=\frac 1 2 c^{+-}_{AA}\hat{S}_B^+,\\ \nonumber
\hat{S}_A^-\hat{S}_B^+\hat{S}_A^+ & \rightarrow & c^{+-}_{AA}\hat{S}_B^+ + \lambda_{AB}c^{+-}_{AB}\hat{S}_A^+,\\ \nonumber
\hat{S}_A^z\hat{S}_B^z\hat{S}_A^+ & \rightarrow & \frac 1 2 \lambda_{AB}c^{+-}_{AB}\hat{S}_A^+,\\ \nonumber
\hat{S}_A^-\hat{S}_B^+\hat{S}_B^+ & \rightarrow & 2\lambda_{AB}c^{+-}_{AB} \hat{S}_B^+,
\end{eqnarray}
where $A\ne B\ne C\ne A$ are sites of the kagome lattice, $c^{+-}_{AB}=\langle\hat{S}_{A}^{+}\hat{S}_{B}^{-}\rangle$
and the conservation of total $S^z$ is implied, i.e.,
$c^{+z}_{AB}=c^{-z}_{AB}=0$.

In Eq.~(\ref{decoup_RGM}) two classes of so-called vertex parameters, 
$\alpha^{}_{AB}$ and $\lambda^{}_{AB}$, 
are introduced to improve the approximation made by the decoupling.
The  parameter $\alpha^{}_{AB}$ 
enters the decoupling scheme if all sites are different from each other,
see lines 1 and 2 in Eq.~(\ref{decoup_RGM}).
In line 3 of Eq.~(\ref{decoup_RGM}) 
the correlation $\langle \hat{S}_{A}^{+}\hat{S}_{A}^{-}\rangle$ is determined 
by using the sum rule (operator identity) $\hat{\mathbf{S}}^2= \hat{S}^+\hat{S}^- - \hat{S}^z +
(\hat{S}^z)^2$, i.e., due to $\langle \hat{S}^z\rangle=0$
within the RGM we have $3\langle (\hat{S}^z)^2\rangle =
\langle \hat{\mathbf{S}}^2 \rangle=  \langle \hat{S}^+\hat{S}^-\rangle
+\langle (\hat{S}^z)^2 \rangle$ and finally $\langle \hat{S}_{A}^{+}\hat{S}_{A}^{-}\rangle = \frac{2}{3}S(S+1)$.
The other class of  
vertex parameters, $\lambda^{}_{AB}$, present in lines 4, 5 and 6 of Eq.~(\ref{decoup_RGM})  
appears only for $S>1/2$ if two sites coincide 
and the remaining correlation function cannot be 
obtained by an operator identity.

Then the EoM reads
\begin{eqnarray}
(\omega^2\mathbb{I}-F_{\mathbf{q}})\chi^{+-}_\mathbf{q}(\omega)&=&-M_{\mathbf{q}}, \label{rgm_green}
\end{eqnarray}
where $M_\mathbf{q}$ (moment matrix), $F_{\mathbf{q}}$ (frequency matrix) and $\chi_\mathbf{q}$ (susceptibility matrix)
are hermitian 3$\times$3-matrices and  $\mathbb{I}$ is the identity matrix. 
Performing corresponding calculations as described above
the components  $M^{\alpha
\beta}_\mathbf{q}=\langle[\textrm{i}\dot{\hat{S}}_{\mathbf{q}\alpha}^{+},\hat{S}_{\mathbf{q}\beta}^{-}]\rangle$ 
of the moment matrix are obtained as
\begin{eqnarray}
M_{\mathbf{q}}= & & \\  4Jc_{1,0} & &\left(\begin{array}{ccc}
-2 & \cos(\frac{\sqrt{3}q_{x}-q_{y}}{2}) & \cos(q_{y})\\ \nonumber
\cos(\frac{\sqrt{3}q_{x}-q_{y}}{2}) & -2 & \cos(\frac{\sqrt{3}q_{x}+q_{y}}{2})\\
\cos(q_{y}) & \cos(\frac{\sqrt{3}q_{x}+q_{y}}{2}) & -2
\end{array}\right). \label{Mq} \\ \nonumber
\end{eqnarray}
The elements of frequency matrix of the spin excitations  
\begin{eqnarray}
 F_{\mathbf{q}}= \left(\begin{array}{ccc} 
F_{\mathbf{q}}^{1,1} & F_{\mathbf{q}}^{1,2} & F_{\mathbf{q}}^{1,3}\\
F_{\mathbf{q}}^{1,2} & F_{\mathbf{q}}^{2,2} & F_{\mathbf{q}}^{2,3}\\
F_{\mathbf{q}}^{1,3} & F_{\mathbf{q}}^{2,3} & F_{\mathbf{q}}^{3,3}
\end{array}\right), \\ \nonumber
\end{eqnarray}
are given by
\begin{eqnarray}
\frac{3}{2}J^{-2}F_{\mathbf{q}}^{1,1}=6\tilde{\lambda}_{1,0}+6\tilde{\alpha}_{1,1}+6\tilde{\alpha}_{2,0}+4S(S+1) \\ 
	+3\left(\cos\left(\sqrt{3}q_{x}-q_{y}\right)+\cos(2q_{y})+2\right)\tilde{\alpha}_{1,0}, \nonumber\\ 
\frac{3}{4}J^{-2}F_{\mathbf{q}}^{2,2}	=3\tilde{\lambda}_{1,0}+3\tilde{\alpha}_{1,1}+3\tilde{\alpha}_{2,0}+2S(S+1) \nonumber\\ 
	+3\left(\cos\left(\sqrt{3}q_{x}\right)\cos(q_{y})+1\right)\tilde{\alpha}_{1,0}, \nonumber\\ 
\frac{3}{2}J^{-2}F_{\mathbf{q}}^{3,3}	=6\tilde{\lambda}_{1,0}+6\tilde{\alpha}_{1,1}+6\tilde{\alpha}_{2,0}+4S(S+1) \nonumber\\ 
	+3\left(\cos\left(\sqrt{3}q_{x}+q_{y}\right)+\cos(2q_{y})+2\right)\tilde{\alpha}_{1,0}, \nonumber 
\end{eqnarray}
\begin{widetext}
\begin{eqnarray}
(\sqrt{2}J)^{-2}F_{\mathbf{q}}^{1,2}&=& \cos\left(\frac{1}{2}\left(\sqrt{3}q_{x}+3q_{y}\right)\right)\tilde{\alpha}_{1,0} \\ 
&-&\cos\left(\frac{1}{2}\left(\sqrt{3}q_{x}-q_{y}\right)\right)\left(\tilde{\lambda}_{1,0}+3\tilde{\alpha}_{1,0}
+\tilde{\alpha}_{1,1}+\tilde{\alpha}_{2,0}+\frac{2}{3}S(S+1)\right), \nonumber\\
(\sqrt{2}J)^{-2}F_{\mathbf{q}}^{1,3}&=& \cos\left(\sqrt{3}q_{x}\right)\tilde{\alpha}_{1,0}-\cos(q_{y})\left(\tilde{\lambda}_{1,0}
+3\tilde{\alpha}_{1,0}+\tilde{\alpha}_{1,1}+\tilde{\alpha}_{2,0}+\frac{2}{3}S(S+1)\right), \nonumber\\ 
(\sqrt{2}J)^{-2}F_{\mathbf{q}}^{2,3}&=& \cos\left(\frac{1}{2}\left(\sqrt{3}q_{x}-3q_{y}\right)\right)\tilde{\alpha}_{1,0}\nonumber \\
&-&\cos\left(\frac{1}{2}\left(\sqrt{3}q_{x}+q_{y}\right)\right)\left(\tilde{\lambda}_{1,0}
+3\tilde{\alpha}_{1,0}+\tilde{\alpha}_{1,1}+\tilde{\alpha}_{2,0}+\frac{2}{3}S(S+1)\right), \nonumber 
\end{eqnarray}
\end{widetext}
where we have used the abbreviations
\begin{eqnarray}
\tilde{\alpha}_{i,j} =  \alpha_{i,j}c_{i,j},
  \quad
  \tilde{\lambda}_{i,j} =  \lambda_{i,j}c_{i,j},
\end{eqnarray}
and  lattice symmetry is used to identify equivalent correlators.
The indices $i,j$  indicate lattice sites
separated by the vector
$\mathbf{R}_{i,j}=\mathbf{r}_i-\mathbf{r}_j=i\mathbf{a_1}/2+j\mathbf{a_2}/2$,
i.e.,
$c_{ij} \equiv \langle \hat{S}^+_{\mathbf{0}}
\hat{S}^-_{\mathbf{R}_{i,j}}\rangle$. 
Their common eigenvectors $|{\gamma\mathbf{q}}\rangle$ and their eigenvalues ($M_{\mathbf{q}}|{\gamma\mathbf{q}}\rangle=m_{\gamma\mathbf{q}}|{\gamma\mathbf{q}}\rangle, F_{\mathbf{q}}|{\gamma\mathbf{q}}\rangle=\omega^2_{\gamma\mathbf{q}}|{\gamma\mathbf{q}}\rangle$, with $\gamma=1,2,3$)
are needed to solve a system of self-consistent equations. The square-root of the eigenvalues
of the frequency matrix $F_\mathbf{q}$ 
 can be identified
as the branches $\omega_{\gamma\mathbf{q}}$, $\gamma=1,2,3$, of the excitation spectrum.

Finally, 
the dynamic wavelength-dependent susceptibility reads 
\begin{equation} \label{chi_oq}
\chi^{+-}_\mathbf{q\alpha\beta}(\omega) 
= -\sum_{\gamma} \frac{m_{\gamma\mathbf{q}}}{\omega^2-\omega^2_{\gamma\mathbf{q}}}\langle\alpha|{\gamma\mathbf{q}}\rangle\langle{\gamma\mathbf{q}}|\beta\rangle
\end{equation}
and the 
static $\mathbf{q}$-dependent susceptibility is given by
\begin{equation} \label{chi_q}
\chi_{\mathbf{q}}= \lim_{\omega\rightarrow0}\frac{1}
{2n_{\rm uc}}\sum_{\alpha,\beta}
\chi^{+-}_{\mathbf{q}\alpha\beta}(\omega), 
\end{equation}
where $n_{\rm uc}=3$ is the number of sites in the geometric unit cell.
The 
correlation functions are obtained by applying the spectral theorem
\begin{eqnarray}
c_{m\alpha,n\beta} & = & \frac{1}{\mathcal{N}}\sum_{\mathbf{q}}c_{\mathbf{q}\alpha\beta}\cos(\mathbf{q}\mathbf{r}_{m\alpha,n\beta}),\label{eq_correlRGM}
\end{eqnarray}
with
\begin{eqnarray}
c_{\mathbf{q}\alpha\beta}&=&\sum_{\gamma}\frac{m_{\gamma\mathbf{q}}}{2\omega_{\gamma\mathbf{q}}}(1+2n(\omega_{\gamma\mathbf{q}}))
\langle\alpha|{\gamma\mathbf{q}}\rangle\langle{\gamma\mathbf{q}}|\beta\rangle,    \label{eq_cqRGM}
\end{eqnarray}
where $\mathcal{N}$ is the number of unit cells and $n(\omega_{\gamma\mathbf{q}})$ is the Bose-Einstein distribution function. 
At the $\Gamma$ point ($\mathbf{q}=\mathbf{0}$) the eigenvectors have the very simple form $|{1\mathbf{0}}\rangle=(1,0,-1)/\sqrt{2}$, $|{2\mathbf{0}}\rangle=(1,-2,1)/\sqrt{6}$,
and $|{3\mathbf{0}}\rangle=(1,1,1)/\sqrt{3}$.

After straightforward calculations we get
\begin{eqnarray} \label{eigsys}
  m_{1{\bf q}} & = & -12J c_{1,0},\\
  \nonumber
  m_{2{\bf q}} & = & -2J c_{1,0} (3 + D_{\bf q}),\\
  \nonumber
  m_{3{\bf q}} & = & -2J c_{1,0}(3 -D_{\bf q}),\\
  \nonumber
  \omega_{1{\bf q}}^2 & = & 6J^2(\frac{2}{3}S(S+1) +   \tilde{\lambda}_{1,0} +  2\tilde{\alpha}_{1,0} + \tilde{\alpha}_{1,1} +  \tilde{\alpha}_{2,0}),\\
  \nonumber
  \omega_{2{\bf q}}^2 & = &  J^2(\frac{2}{3}S(S+1) + \tilde{\lambda}_{1,0} + 2\tilde{\alpha}_{1,0} + \tilde{\alpha}_{1,1} + \tilde{\alpha}_{2,0} \\  \nonumber
  &  & - \tilde{\alpha}_{1,0} (3 - D_{\bf q}))(3 + D_{\bf q}),\\
  \nonumber
  \omega_{3{\bf q}}^2 & = &  J^2(\frac{2}{3}S(S+1) + \tilde{\lambda}_{1,0} + 2\tilde{\alpha}_{1,0} + \tilde{\alpha}_{1,1} + \tilde{\alpha}_{2,0} \\ \nonumber
  &  & - \tilde{\alpha}_{1,0} (3 + D_{\bf q}))(3-D_{\bf q}),\\
  \nonumber
  D_{\bf q}^2 & = & 3+2\cos(2q_y)+2\cos(\sqrt{3}q_x-q_y)\\
  \nonumber
   & &+2\cos(\sqrt{3}q_x+q_y) .
\end{eqnarray}
Obviously, we have one flat band, namely $\omega_{1{\bf q}}$, and two dispersive branches  $\omega_{2{\bf q}}$ and
 $\omega_{3{\bf q}}$, where $\omega_{3{\bf q}}$ is the acoustic branch.

The static uniform susceptibility is given by (cf. Eqs.~(\ref{chi_oq}) and (\ref{chi_q}))
\begin{eqnarray}
\chi_{0}&=&\lim_{\mathbf{q}\rightarrow \mathbf{0}}\chi_{\mathbf{q}}\label{eq_susc0RGM} 
=\underset{\mathbf{q}\rightarrow\mathbf{0}}{\textrm{lim}}\frac{m_{3\mathbf{q}}}{2\omega^2_{3\mathbf{q}}} \\
&=&\frac{-c_{1,0}}{J(\frac{2}{3}S(S+1)+\tilde{\lambda}_{1,0}-4\tilde{\alpha}_{1,0}+\tilde{\alpha}_{1,1}+\tilde{\alpha}_{2,0})}. \nonumber
\end{eqnarray}
The magnetic correlation length $\xi_{\mathbf{Q}}$ is obtained by
expanding the susceptibility $\chi_{\mathbf{Q+q}}=\sum_{\alpha,\beta}\chi^{+-}_{\alpha\beta\mathbf{Q+q}}/(2n_{uc})\approx\chi_{\mathbf{Q}}/(1+\xi_{\mathbf{Q}}^2\mathbf{q}^2)$
in the neighborhood of the corresponding magnetic wave vector $\mathbf{Q}$, see, e.g., \cite{Schmalfus2004,Schmalfus2005,Junger2005,Haertel2010,Haertel2011,Haertel2011a,Haertel2013,Mueller2015,2017Mullera,2017Muller}.
While for the $q=0$ state the expansion is straightforward and yields $\xi_{\mathbf{Q}_0}=\sqrt{J\alpha_{1,0}\chi_{\mathbf{Q}_0}}$, the corresponding
susceptibility for the $\sqrt{3}\times \sqrt{3}$ state $\chi_{\mathbf{Q}_1}
=\frac{-c_{1,0}}{J(\frac{2}{3}S(S+1)+\tilde{\lambda}_{1,0}+2\tilde{\alpha}_{1,0}+\tilde{\alpha}_{1,1}+\tilde{\alpha}_{2,0})}=m_{1{\bf q}}/(2\omega_{1{\bf
q}}^2)$ is a quotient of two $\mathbf{q}$-independent quantities, cf.
Eq.~(\ref{eigsys}).
Having in mind the above relation between $\xi_{\mathbf{Q}_0}$ and
$\chi_{\mathbf{Q}_0}$ and the fact that both quantities would simultaneously diverge at a
transition point to
magnetic LRO, see, e.g., \cite{2017Mullera,2017Muller}, we choose $\xi_{\mathbf{Q}_1}=\sqrt{J\alpha_{1,0}\chi_{\mathbf{Q}_1}}$
as a measure of the correlation length related to a possible $\sqrt{3}\times \sqrt{3}$
ordering. In what follows we will use the term 'correlation
length' for $\xi_{\mathbf{Q}_1}$, too.
To analyze magnetic ordering we can use the 
static magnetic structure factor ${\cal S}(\mathbf{q})=(1/N)\sum_{i,j}\langle
\hat{\mathbf{S}}_i\hat{\mathbf{S}}_j\rangle\cos(\mathbf{q}\mathbf{R}_{i,j})$, which is
related to 
$c_{\mathbf{q}\alpha\beta}$, cf. Eq.~(\ref{eq_cqRGM}).

The final step in the RGM approach is to find as many equations as there are unknown
quantities in the RGM equations,
where except the correlation functions entering the EoM also the introduced vertex parameters  $\alpha_{i,j}(T)$ and
$\lambda_{i,j}(T)$ have to be determined.
Then, by numerical solution of the resulting system of
coupled
self-consistent equations the physical quantities can be determined.
Taking into account all possible vertex parameters $\alpha_{i,j}(T)$ and $\lambda_{i,j}(T)$
would noticeably exceed the number of available equations.
Within the minimal version of the RGM one takes into account only one vertex parameter
in  each class,
i.e., $\alpha_{i,j}(T)=\alpha(T)$ and $\lambda_{i,j}(T)=\lambda(T)$. Note
that this simple version with only one $\alpha$ parameter ($\lambda(T)\equiv0$) was used in the
early RGM kagome papers 
for the spin-half case, see \cite{Yu2000,Bernhard2002,Schmalfus2004}.
This  
approach is particularly appropriate for ferromagnets
\cite{Kondo1972,Suzuki1994,Haertel2008,Junger2005,Schmalfus2005,Antsygina2008,Haertel2010,Haertel2011,Haertel2011,Mueller2015,2017Mullera,2017Muller}, 
where all correlation
functions have the same sign.
However, for antiferromagnets typically the consideration of one additional vertex
parameter 
allowing to distinguish between nearest-neighbor and
further-neighbor correlations may yield a significant improvement of the
method, see, e.g.,
\cite{Ihle1997,Ihle2001,Schmalfus2006,Haertel2013,Vladimirov2014,Vladimirov2017}.
Thus, we set
$\alpha_{i,j}(T)=\alpha_{1}(T)$, if ($i,j$) are nearest neighbors sites, 
and $\alpha_{i,j}(T)=\alpha_{2}(T)$, if ($i,j$) are not nearest neighbors
sites.  (For the minimal version $\alpha_{2}=\alpha_{1}$
holds.)
Note that  in the
relevant equations the vertex parameters $\lambda_{i,j}(T)$ only appear for  nearest-neighbors sites $i$ and $j$,
i.e., we set
consistently  $\lambda_{i,j}(T)=\lambda(T)$.

The required equations to determine all unknown quantities are as follows:
For every unknown correlation function the spectral theorem yields one
equation, cf.  Eqs.~(\ref{eq_correlRGM}) and (\ref{eq_cqRGM}). 
Another equation is given by the sum rule $\bm{\hat{S}}_{m\alpha}^2 =S(S+1)$,
 which determines, e.g., one vertex parameter, say $\alpha_1$. 
For the missing two vertex parameters, $\alpha_2$  and $\lambda$,
we follow
\cite{Shimahara1991,Ihle1997,Ihle2001,Junger2005,Juh_sz_Junger_2009,Haertel2011,Vladimirov2014,Mueller2015,2017Muller,Vladimirov2017} 
and use the ansatzes 
$r_1(T)=(\alpha_{1}(T)-\alpha_{1}(\infty))/(\lambda(T)-\lambda(\infty))=r_1(0)$ and 
 $r_2(T)=(\alpha_{1}(T)-\alpha_{1}(\infty))/(\alpha_{2}(T)-\alpha_{2}(\infty))=r_2(0)$, where
 the values
$\alpha_1(\infty)=\alpha_2(\infty)=1$ and $\lambda(\infty)=1-3/(4S(S+1))$  are known 
and can be verified by comparison with the high-temperature expansion, 
see, e.g., \cite{Junger2005}.
(Note that in the minimal version of the RGM only one of these two equations, namely $r_1(T)$, has to be solved, 
because $\alpha_{1}=\alpha_{2}$.)
For the vertex parameter $\lambda(T)$ at zero temperature we use
the well-tested ansatz $\lambda(0)=2-1/S$
\cite{Juh_sz_Junger_2009,Haertel2011,Vladimirov2014,Mueller2015}.
Last but not least, for the extended version we determine the additional vertex
parameter $\alpha_{2}(0)$ 
 by adjusting the GS energy to the values obtained by high-order coupled
 cluster method (CCM) \cite{Goetze2011,Goetze2015}, which is known to yield precise
 values for $E_0$, see, e.g., Fig.~7 in \cite{Xie2014}.

\subsection{High Temperature Expansion (HTE) \label{HTE}}

In addition to the RGM, we use a general high temperature expansion (HTE)
code,
see \cite{Lohmann2011,Lohmann2014}, to discuss the thermodynamics
of the 
KHAF. 
We compute the series of the susceptibility $\chi_{0}=\sum_nc_n\beta^n$ and the specific heat
$C=\sum_nd_n\beta^n$ up
to order 11. To extend the region of validity of the power series Pad\'{e} approximants are a conventional transformation.
These approximants are ratios of two polynomials of degree $m$ and $n$: $[m,n]=P_m(x)/Q_n(x)$. 
Furthermore the series of the correlation functions $\langle
\hat{\mathbf{S}}_i
\hat{\mathbf{S}}_j\rangle$ are 
analyzed up to 11th order, which we use
to consider the static magnetic structure factor 
${\cal S}(\mathbf{q})=(1/N)\sum_{i,j}\langle
\hat{\mathbf{S}}_i\hat{\mathbf{S}}_j\rangle\cos(\mathbf{q}\mathbf{R}_{i,j})$,
see, e.g., \cite{Richter2015}.
The structure factor is one of the main outcomes of neutron diffraction
measurements, where the  maxima of the structure factor indicate the favored magnetic
ordering.

\section{Results\label{results}}
In what follows we set the energy scale of the model (\ref{model}) by fixing the
exchange
constant $J=1$.

\subsection{Zero-temperature properties\label{sec:GS}}

\begin{figure}
\centering 
\includegraphics[scale=0.33,angle=-90]{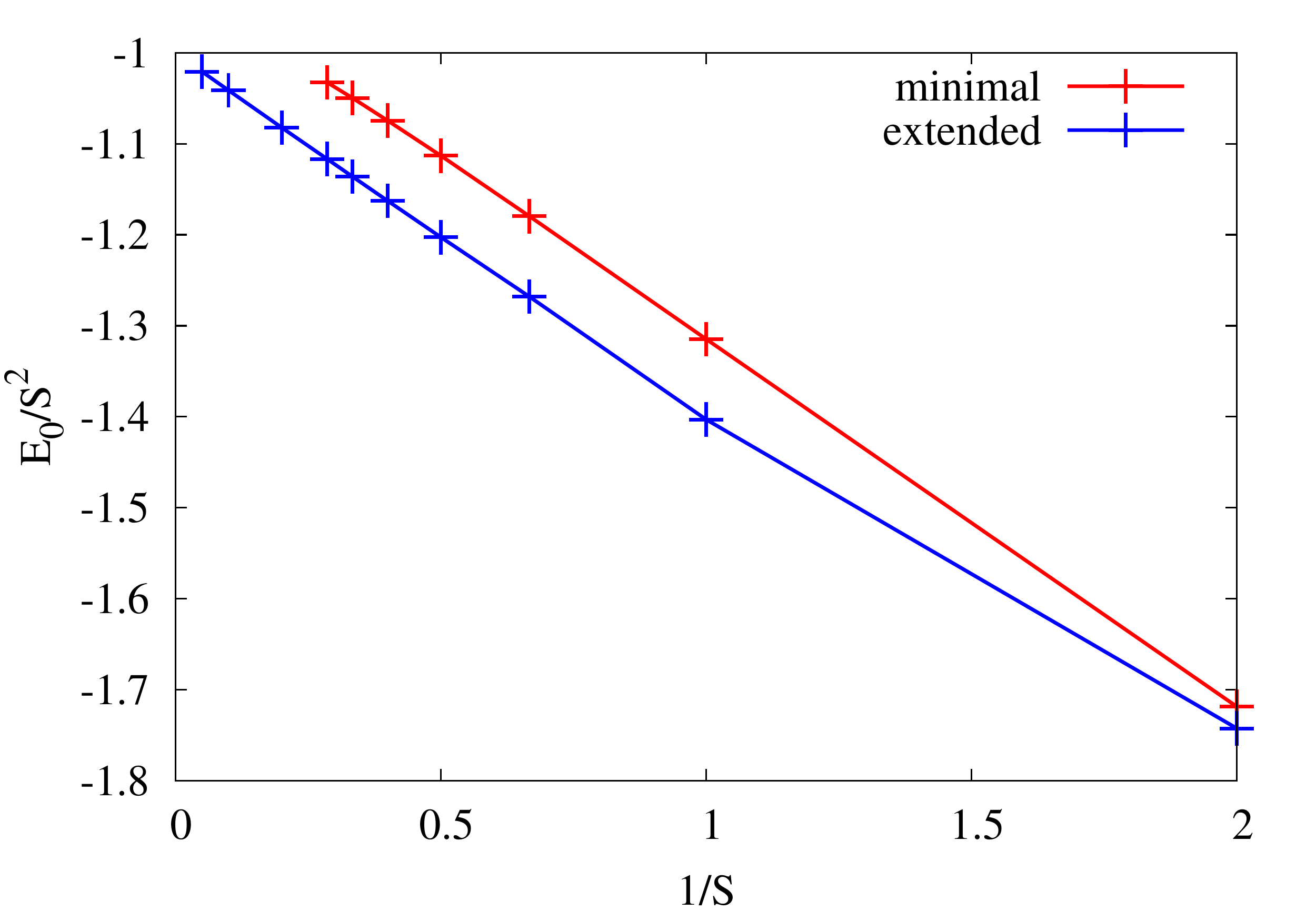} 
\protect\caption{
RGM GS energy per site $E_0/S^2$ in dependence on the inverse spin quantum
number $1/S$  (minimal vs. extended version). Note that $E_0$ for the extended
version naturally coincides with the CCM data of \cite{Goetze2011,Goetze2015}. 
}
\label{Fig2} 
\end{figure}
\begin{figure}
\centering 
\includegraphics[scale=0.33,angle=-90]{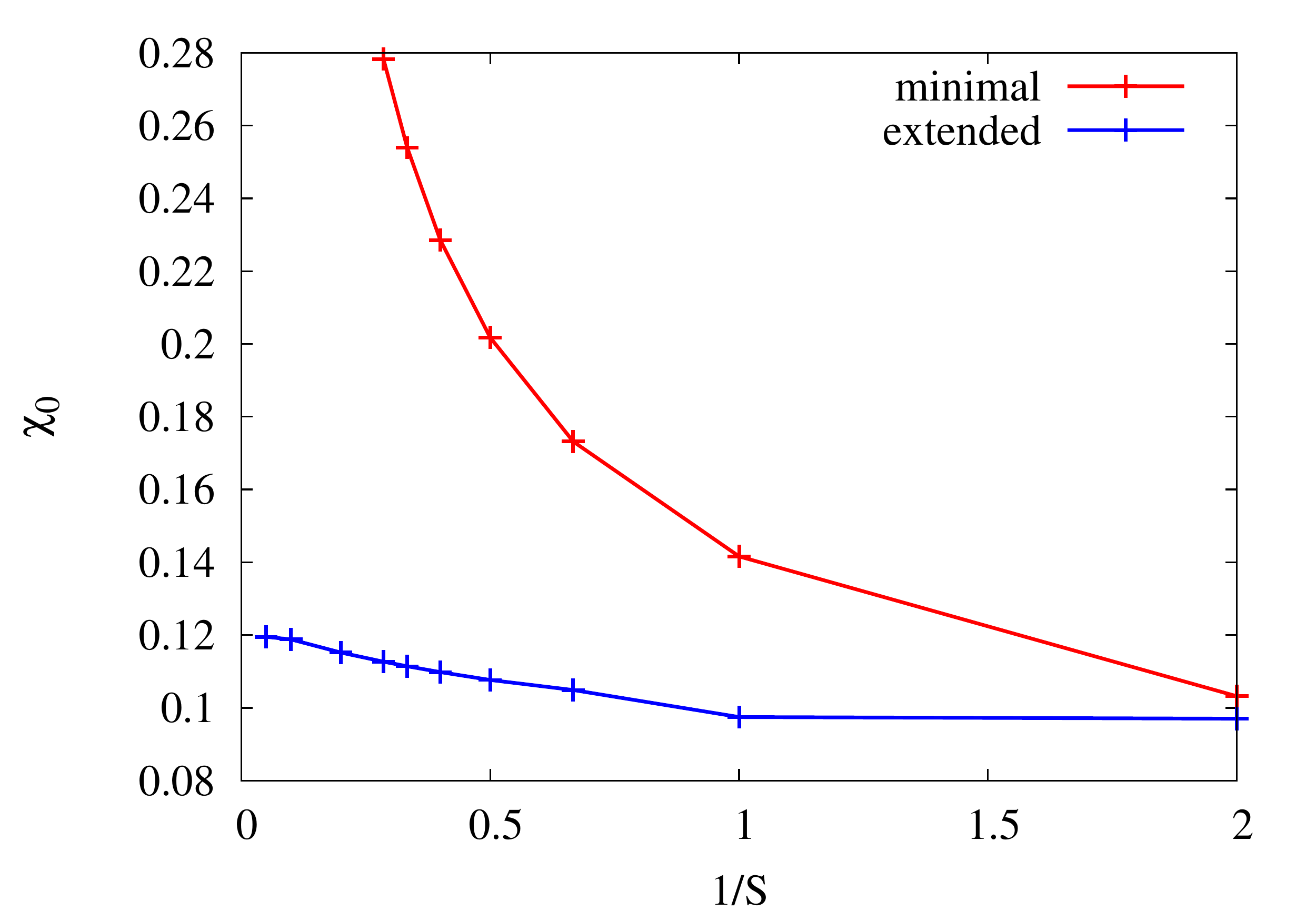} 
\protect\caption{
RGM GS static uniform susceptibility $\chi_{0}$ in dependence on the inverse
spin quantum
number $1/S$ (minimal vs. extended version).
}
\label{Fig3} 
\end{figure}

\begin{figure}
\centering 
\includegraphics[scale=0.85]{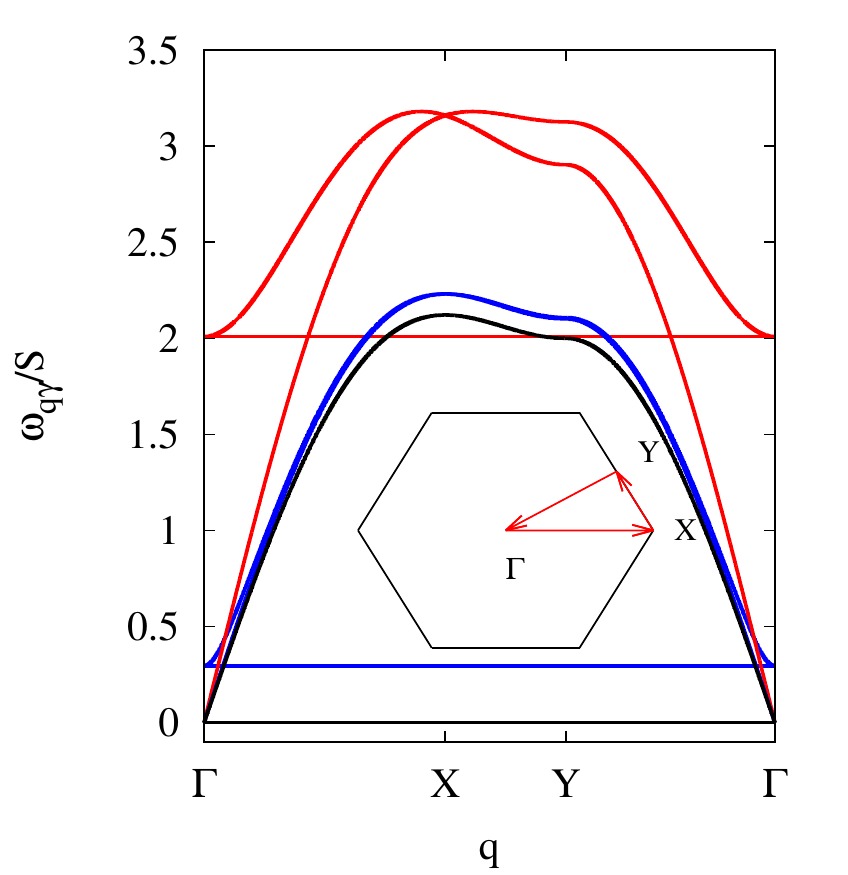} 
\protect\caption{
RGM GS results for the dispersion of the magnetic excitations $\omega_{\gamma\mathbf{q}}/S$ ($\gamma=1,2,3$) 
for $S=1/2$ (red lines) and $S=3$ (blue lines) compared with data of the
LSWT
(black lines)  along a typical path in the first Brillouin zone
(see inset).
LSWT formulas for $\omega_{\gamma\mathbf{q}}/S$  can be found, e.g., in
\cite{Chernyshev2015}.
}
\label{Fig4} 
\end{figure}

\begin{figure}
\centering 
\includegraphics[scale=0.33,angle=0]{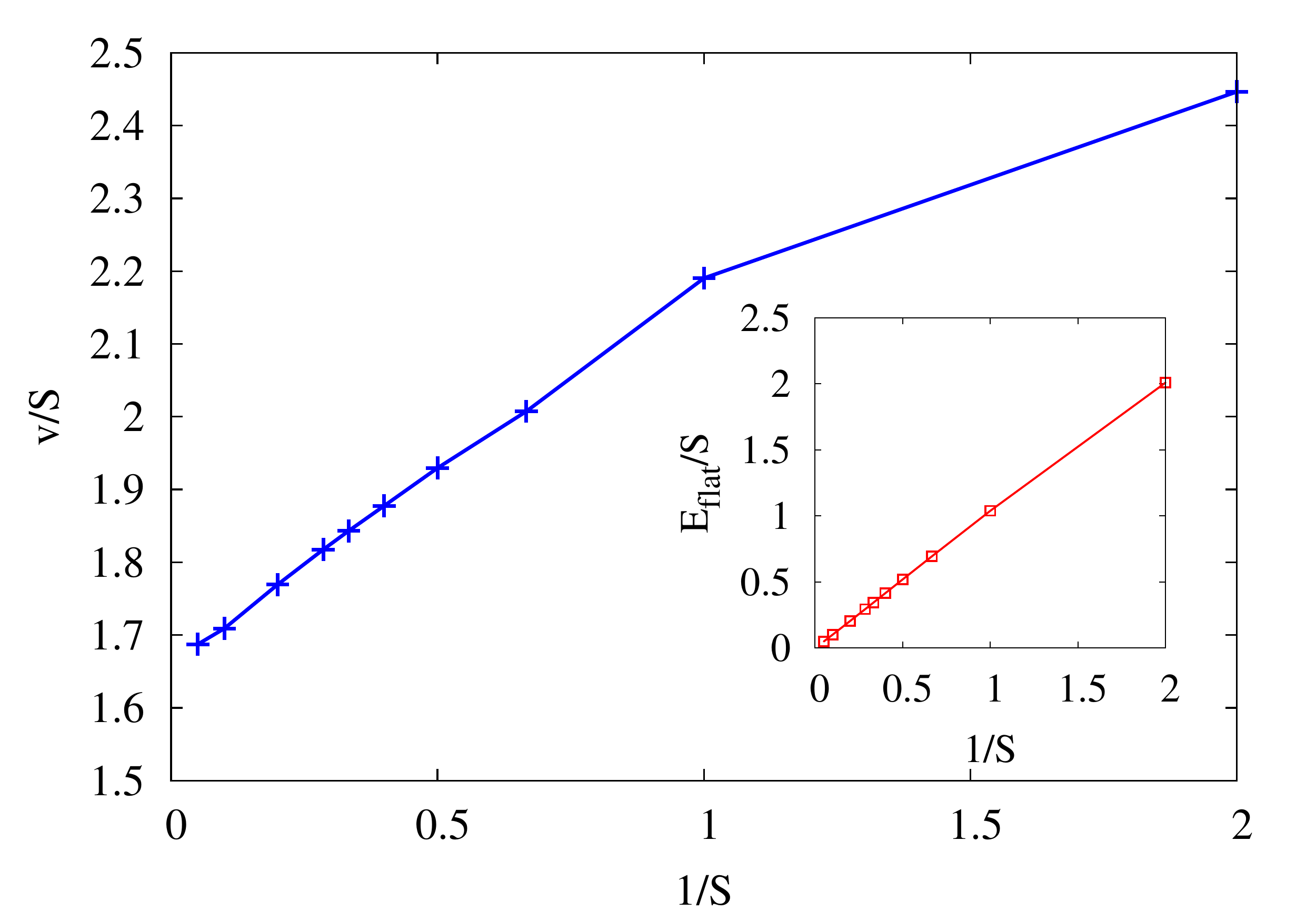} 
\protect\caption{
Main: Normalized RGM GS excitation velocity $v(0)/S$ 
in dependence on the inverse spin quantum
number $1/S$. Inset: Position $E_{\rm flat}/S$ of the flat band
in dependence on the inverse spin quantum
number $1/S$.
}
\label{Fig5} 
\end{figure}

We start with the discussion of the RGM results for the GS
properties using the minimal as well as the extended (i.e., with CCM input) version
of the RGM.
In Fig.~\ref{Fig2} we show  the GS energy $E_0/S^2$ as a function of
the inverse spin quantum number $1/S$.
It is obvious that the minimal version leads to significant higher energy
values, where for $S=1/2$ the difference is smallest.
It is also obvious, that the minimal version does not yield the correct classical large-$S$ limit, $\lim_{S\to\infty}
E_0/S^2 = -1$. Thus, we conclude that the minimal version is only applicable
for small values of $S$. This conclusion  is supported by the data for the
static uniform susceptibility $\chi_{0}$ shown in Fig.~\ref{Fig3}.
In what follows (i.e., figures subsequent to
Fig.~\ref{Fig3}), we therefore focus on the discussion of the results obtained
by the extended version, i.e., unless stated otherwise, all data presented
below
belong to the extended version.
Now we discuss the excitation spectrum shown in Fig.~\ref{Fig4}. 
We mention first, that in linear spin-wave theory (LSWT)
$\omega_{\gamma\mathbf{q}}/S$ is independent of $S$, the flat band
$\omega_1$  is exactly
at zero energy and the two dispersive branches,
$\omega_{2\mathbf{q}}$ and $ \omega_{3\mathbf{q}}$, are degenerate
\cite{Chernyshev2015}.
The RGM provides an improved description of the excitation energies.  The flat band is of course also present, but its position
$E_{\rm flat}$
depends on $S$, where $E_{\rm flat}/S$ decreases almost linearly with $1/S$ down to
$E_{\rm flat}/S = 0$ as $S \to \infty$,
see inset of Fig.~\ref{Fig5}. Moreover,
the degeneracy of $\omega_{2\mathbf{q}}$ and $ \omega_{3\mathbf{q}}$ is
lifted and 
there is a noticeable dependence of the dispersive branches on $S$.
In particular, in the extreme quantum case $S=1/2$ the dispersion
relations deviate strongly from the LSWT.
As increasing $S$ the RGM data approach the LSWT result.
    
The GS excitation velocity $v$ corresponding to the linear expansion of the
lowest branch  $\omega_{3\mathbf{q}}$ around the
$\Gamma$ point is given by  
$v^2=(\frac{2}{3}S(S+1) + \tilde{\lambda}_{1,0} -4\tilde{\alpha}_{1,0} + \tilde{\alpha}_{1,1} + \tilde{\alpha}_{2,0})$.
The LSWT result is $v_{LSWT}=\sqrt{3} S$.
Numerical data for $v$ are   
shown in Fig.~\ref{Fig5}. While in LSWT  $v/S$ is
 independent of $S$, within the RGM  there is a
noticeable dependence of  $v/S$ on $S$. 

\begin{figure}
\centering 
\includegraphics[scale=0.33,angle=0]{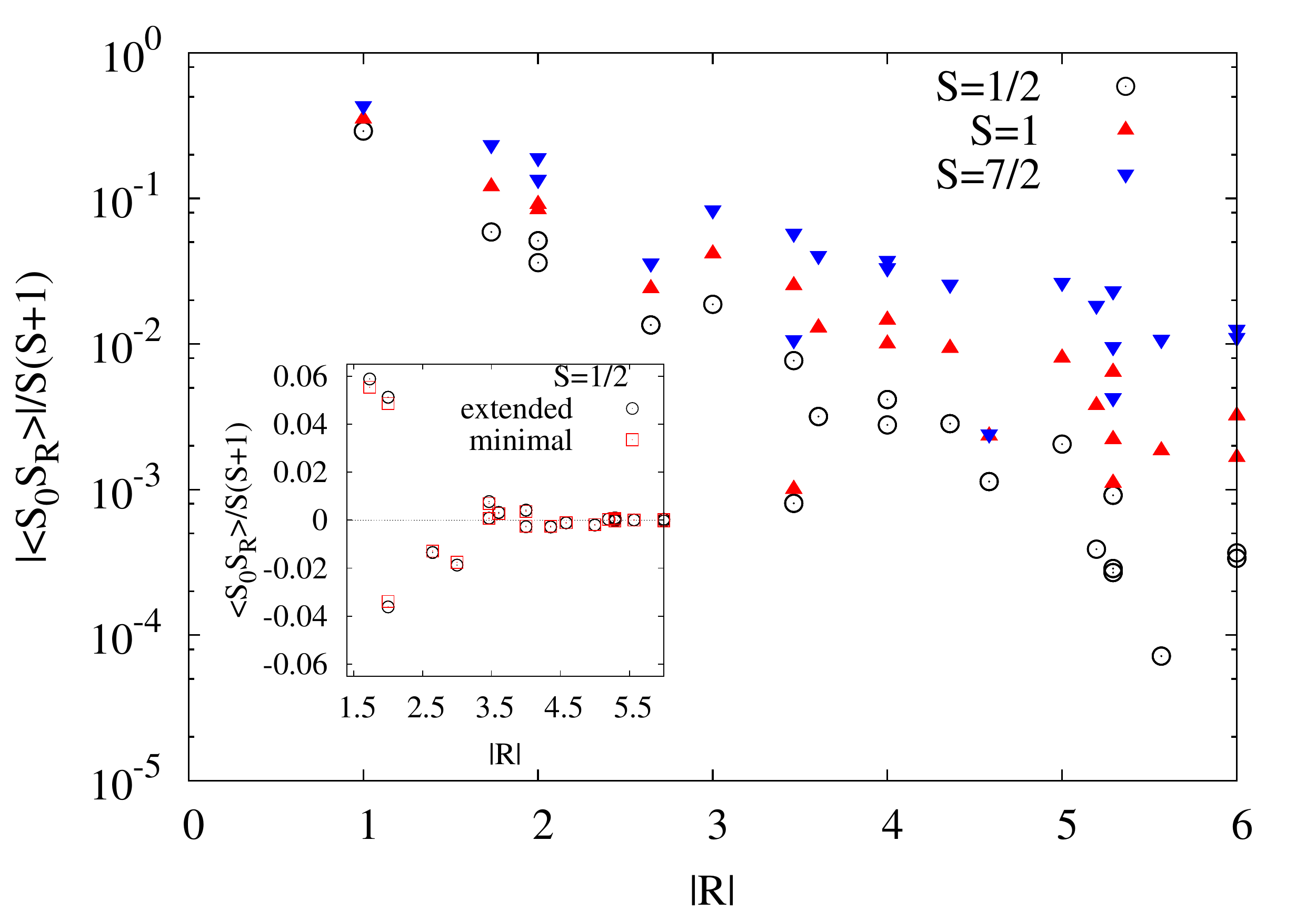} 
\protect\caption{
Main: Magnitude of the GS correlation functions $|\langle
\hat{\mathbf{S}}_0 \hat{\mathbf{S}}_\mathbf{R}\rangle|/S(S+1)$ within
a range of separation
$|\mathbf{R}| \le 6$ for spin  quantum numbers $S=1/2,1$ and $7/2$.
Inset: Comparison of $\langle \hat{\mathbf{S}}_0 \hat{\mathbf{S}}_\mathbf{R}\rangle/S(S+1)$ of the minimal and
extended version of the RGM for $S=1/2$ (nearest-neighbor 
correlation not included).  
}
\label{Fig6} 
\end{figure}

\begin{figure}[ht!]
\centering 
\includegraphics[scale=0.75,angle=0]{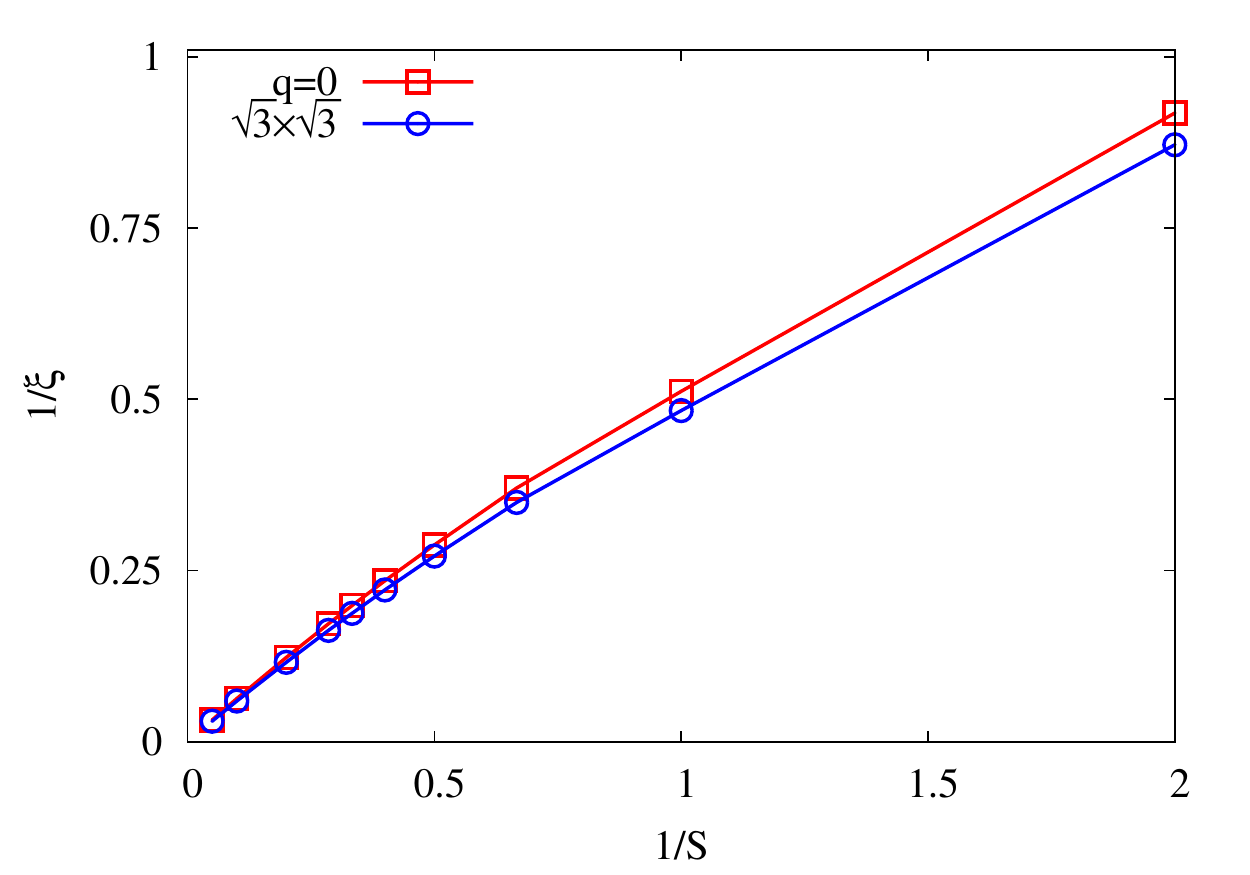} 
\protect\caption{RGM GS correlation lengths corresponding to $q=0$ ($\xi_{\mathbf{Q}_0}$)
and $\sqrt{3}\times \sqrt{3}$
($\xi_{\mathbf{Q}_1}$) ordering.
}
\label{Fig7} 
\end{figure}
\begin{figure*}[ht!]
\centering \includegraphics[scale=0.85]{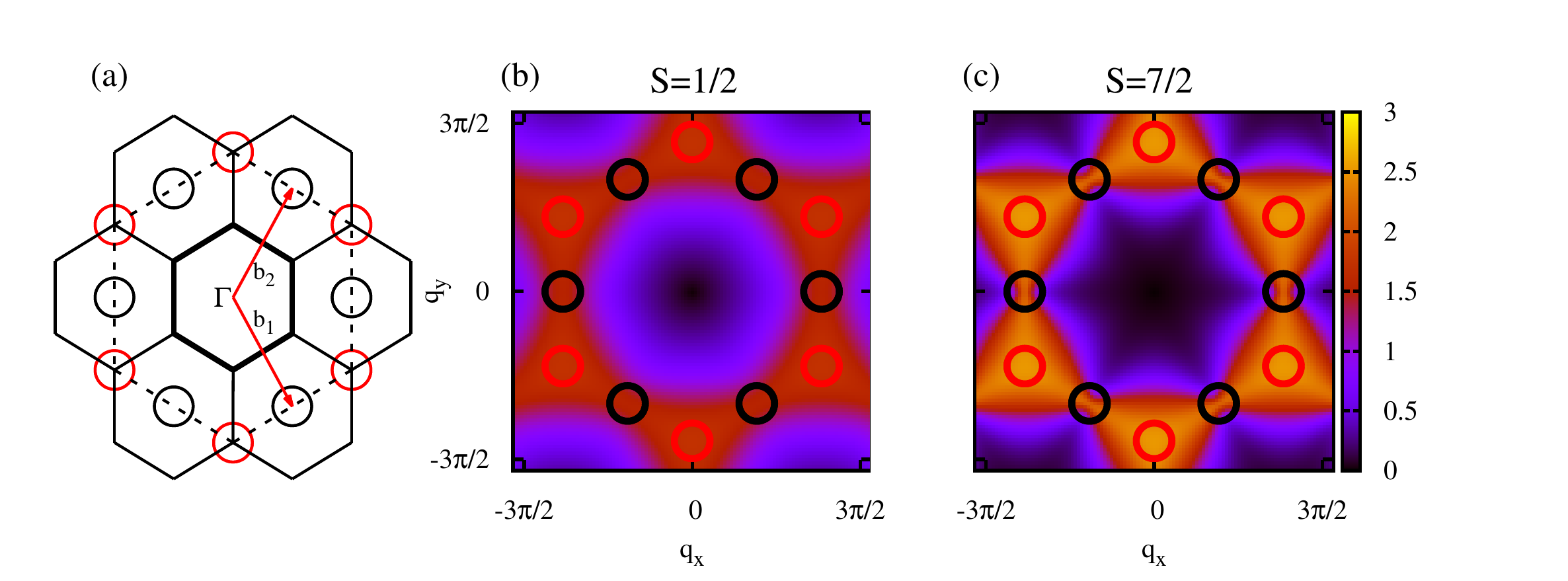} \protect\caption{
RGM GS structure factor ${\cal S}(\mathbf{q})/S(S+1)$. (a) 
Brillouin zones: the solid and dashed lines show the first and
extended Brillouin zones; the red (black) circles indicate the expected maxima for a
classical $\sqrt{3}\times \sqrt{3}$ ($q=0$) state.
(b) $S=1/2$,  (c) $S=7/2$.
}
\label{Fig8} 
\end{figure*}

\begin{figure}[ht!]
\centering \includegraphics[scale=0.7]{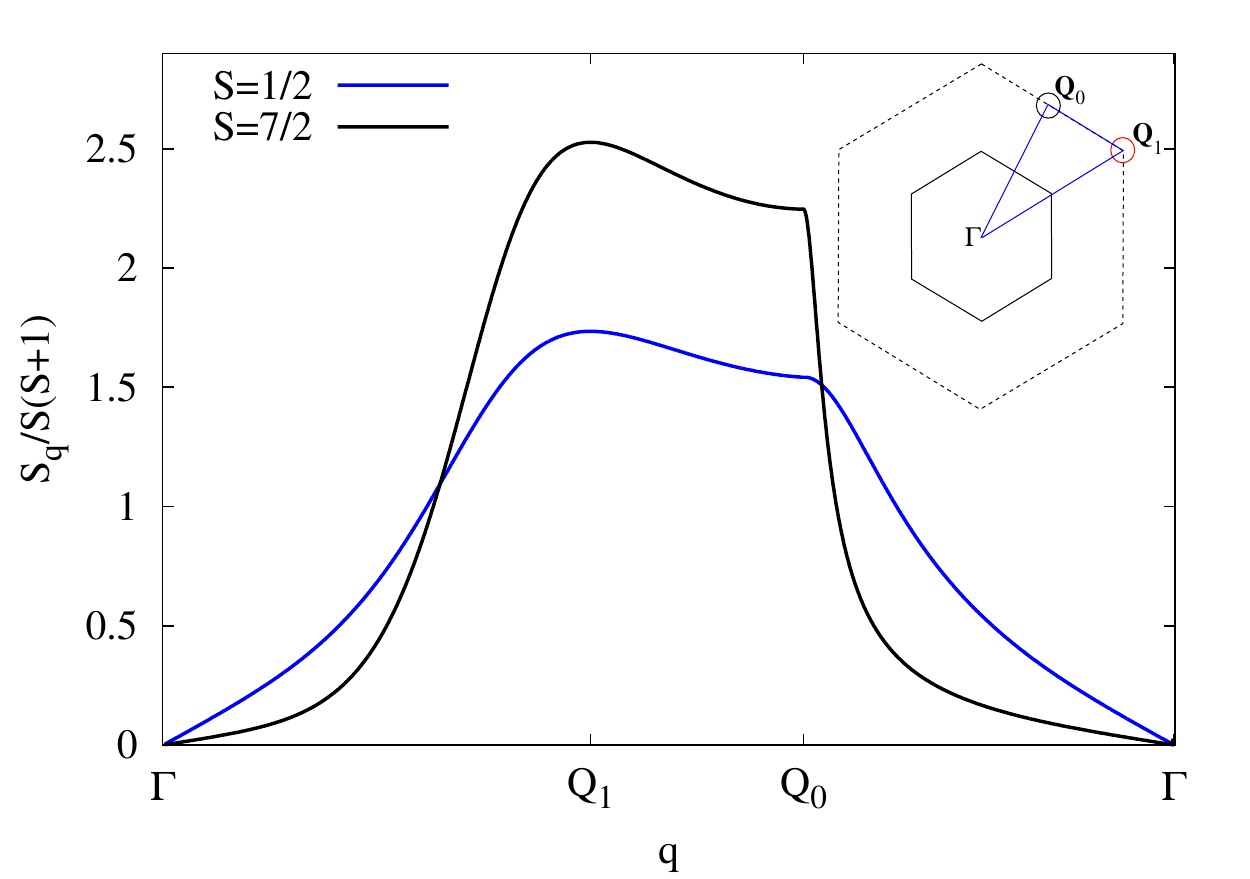} \protect\caption{
Normalized RGM GS 
structure factor ${\cal S}(\mathbf{q})/S(S+1)$ along the path $\Gamma \to
{\mathbf{Q}_1} \to {\mathbf{Q}_0}
\to \Gamma $ 
(see inset) for  $S=1/2$ and $S=7/2$.
}
\label{Fig9} 
\end{figure}

Let us turn to the spin-spin correlation functions $\langle
\hat{\mathbf{S}}_0 \hat{\mathbf{S}}_\mathbf{R}\rangle$. In Fig.~\ref{Fig6}, main panel,  we
show  all non-equivalent GS correlators $\langle \hat{\mathbf{S}}_0
\hat{\mathbf{S}}_\mathbf{R}\rangle/S(S+1)$
up to a separation
$R = |\mathbf{R}|= 6$ for some selected values of $S$ using a logarithmic scale for
$|\langle \hat{\mathbf{S}}_0 \hat{\mathbf{S}}_\mathbf{R}\rangle|/S(S+1)$.
In the inset we compare the minimal with the extended version for $S=1/2$
without using a logarithmic scale.
Note that the presented data for the minimal version correspond to the results
of Bernhard, Canals and Lacroix \cite{Bernhard2002}. 
In accordance with Figs.~\ref{Fig2} and \ref{Fig3} for $S=1/2$ the difference between the minimal and
the extended version are not tremendous but noticeable. 
Since for a certain separation $|\mathbf{R}|$ non-equivalent sites exist, 
more than one data point can appear at one and the same separation $|\mathbf{R}|$.
The data suggest that the overall decay 
of $\ln |\langle \hat{\mathbf{S}}_0 \hat{\mathbf{S}}_\mathbf{R}\rangle/S(S+1)|$ seems to be linear,
thus indicating an exponential decay of the correlators.
It is also obvious, that the decay is faster the lower the spin quantum
number $S$.
This observation from Fig.~\ref{Fig6} is in agreement with results for the correlation lengths $\xi_{\mathbf{Q}_0}$
(corresponding to $q=0$ ordering) 
 and $\xi_{\mathbf{Q}_1}$  (corresponding to 
$\sqrt{3}\times \sqrt{3}$ ordering), shown in Fig.~\ref{Fig7} (for 
 the definition of $\xi_{\mathbf{Q}_0}$
 and $\xi_{\mathbf{Q}_1}$ see
Sec.~\ref{RGM}). 
In the extreme quantum spin-half case the correlation lengths are of the order of one
lattice spacing as expected in
a spin liquid. That is in agreement with known results, e.g., obtained by large-scale
density-matrix renormalization-group (DMRG) studies \cite{Kolley2015}.
The RGM data then indicate a power-law increase of both,  $\xi_{\mathbf{Q}_0}$  and
$\xi_{\mathbf{Q}_1}$, with increasing $S$, see Fig.~\ref{Fig7}. We find  $\xi_{\mathbf{Q}_1} >
\xi_{\mathbf{Q}_0}$
for all
$S$, but the difference of both
correlation lengths is small.

Now we discuss the GS static magnetic structure
factor ${\cal S}(\mathbf{q})$. In Fig.~\ref{Fig8}  we show an intensity plot of 
${\cal S}(\mathbf{q})/S(S+1)$
using an extended Brillouin zone,
see panel (a) and 
cf. also \cite{Kolley2015}.
For $S=1/2$ we find the typical pattern \cite{Laeuchli2009,Depenbrock2012,Kolley2015},
i.e.,
the intensity is concentrated
along the edge of the extended Brillouin zone, where
${\cal S}(\mathbf{q})$ remains small even at the magnetic $\mathbf{q}$-vectors $\mathbf{Q}_0$ and  $\mathbf{Q}_1$
related to the  $q=0$ and $\sqrt{3}\times \sqrt{3}$ states.
This smooth shape of ${\cal S}(\mathbf{q})$ is related to the fast decay of the
spin-spin correlations, see Fig.~\ref{Fig6}. 
As increasing $S$ the structure
factor develops a more pronounced shape, and pinch points, typical for the
classical KHAF \cite{Zhitomirsky2008}, emerge between
triangular shaped areas of large intensity, see Fig.~\ref{Fig8}c.
This observation is also obvious from Fig.~\ref{Fig9}, where we show the  structure
factor along a prominent path in the extended Brillouin zone.
As indicated by Figs.~\ref{Fig8} and \ref{Fig9}, 
we find that for all values of $S$ the relation ${\cal S}(\mathbf{Q}_1)>{\cal S}(\mathbf{Q}_0)$ holds.
Together with the data for the correlation lengths $\xi_{\mathbf{Q}_0}$ and
$\xi_{\mathbf{Q}_1}$ (Fig.~\ref{Fig7}) we may conclude that $\sqrt{3}\times
\sqrt{3}$ SRO is favored in agreement with previous
investigations \cite{Chubukov:92,Sachdev_1992,Henley:1995,Goetze2011,Zhito_PRL_XXZ}.   

From the static GS properties reported above we conclude that, although the
magnetic SRO with $\sqrt{3}\times \sqrt{3}$ symmetry becomes more and more
pronounced with increasing $S$, within the RGM
approach  
no magnetic LRO for the spin-$S$ KHAF is found. 
We may compare this finding with known GS results obtained by other methods.
Note, however, that for $S>1$ data to compare with are extremely rare.
We mention first that within the LSWT the quantum correction of the sublattice
magnetization always diverges due to the zero-energy flat band, see, e.g.,
\cite{Zhito_PRL_XXZ}. 
As briefly discussed in the introduction, more sophisticated GS methods such as the
CCM and the DMRG yield
evidence  
that for $S=1$ 
semiclassical magnetic LRO is also lacking \cite{Goetze2011,Goetze2015,Changlani2015,Liu2015,Picot2015,Nishimoto2015}.
On the other hand, recent results obtained by CCM, tensor network approaches, and
series expansion 
indicate weak 
GS $\sqrt{3}\times \sqrt{3}$  LRO for $S=3/2$ \cite{Goetze2011,Goetze2015,Oitmaa2016,Liu2016}.
Previous experience in applying the RGM on frustrated
quantum antiferromagnets, see, e.g.,
\cite{Barabanov94,Ihle2001,Schmalfus2004,Haertel2013,Mikheyenkov2013} and references
therein,
indicate, however, that the implementation of rotational invariance by
setting $\langle \hat{S}^z_{i}\rangle=0$ in the equations of
motions may overestimate the tendency to melt semiclassical GS
magnetic LRO in RGM calculations.

\subsection{Finite-temperature properties} 
\label{finite_T}

In what follows, as a rule  we will present the temperature
dependence of physical quantities  using a normalized temperature
$T/S(S+1)$. This choice ensures a spin-independent behavior of the physical
quantities at large temperatures \cite{Lohmann2011}.
Moreover, we mention again that (unless stated otherwise) we present RGM data for the extended version using CCM
input (see above).

\subsubsection{Spin-spin correlation functions, specific heat and uniform
susceptibility}
\label{subsec_S0SR}

\begin{figure}[ht]
\centering \includegraphics[scale=0.33,angle=0]{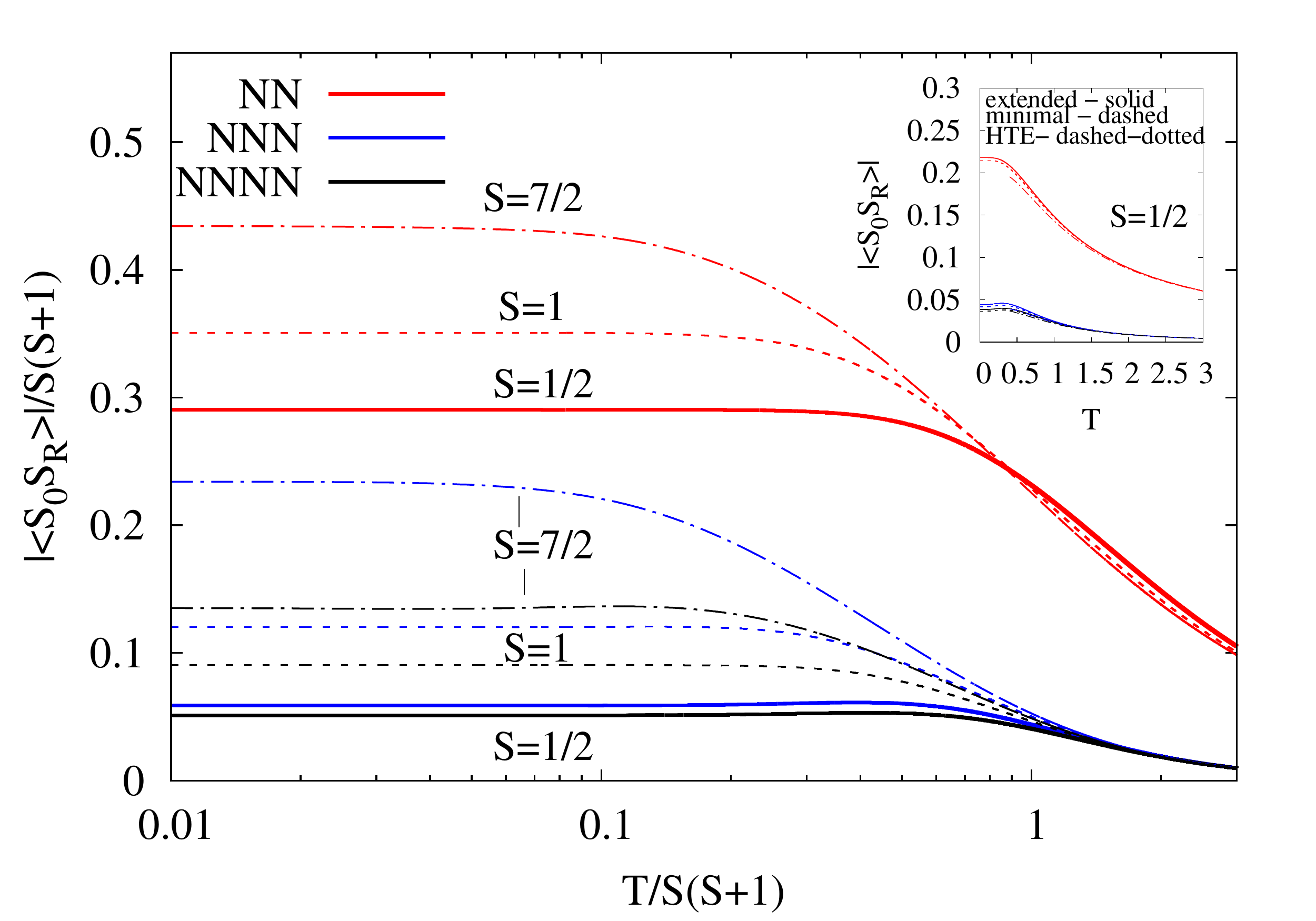} 
\protect\caption{Main panel: Magnitude of the normalized spin-spin correlation functions
$|\langle
\hat{\mathbf{S}}_0 \hat{\mathbf{S}}_\mathbf{R}\rangle|/S(S+1)$  as a function of the normalized temperature
$T/S(S+1)$ (logarithmic scale) 
for spin quantum numbers  $S=1/2, 1$, and $7/2$ (NN -- nearest neighbors;
NNN -- next-nearest neighbors; NNNN -- next-next-nearest neighbors along two
$J_1$ bonds).  
Inset: Magnitude of the  spin-spin correlation functions $|\langle
\hat{\mathbf{S}}_0 \hat{\mathbf{S}}_\mathbf{R}\rangle|$ for $S=1/2$ as a function of the temperature
$T$ (linear scale):  Comparison of the  extended (solid) and minimal
(dashed)
versions of the RGM as well as the 11th-order HTE with subsequent Pad\'e
(dashed-dotted).
}
\label{Fig10} 
\end{figure}

We start with the discussion of the temperature dependence of short-range  spin-spin correlation functions  
$\langle \hat{\mathbf{S}}_0 \hat{\mathbf{S}}_\mathbf{R}\rangle$. We show the absolute values  
in Fig.~\ref{Fig10}, main panel,  for $S=1/2$, $1$, and $7/2$.  (Note that the NN
correlation is antiferromagnetic, whereas the NNN and NNNN  correlation
functions are ferromagnetic.)
We find that there is a low-temperature region
$T/S(S+1) \lesssim 0.1$ where 
the presented  correlation functions are almost temperature independent.
This region  is largest for the extreme quantum case $S=1/2$.
It is also obvious that for $T/S(S+1) < 1$ the magnetic
SRO becomes more pronounced as
increasing $S$ (cf. also Fig.~\ref{Fig6}).
On the other hand, for $T/S(S+1) > 1$ the curves for various $S$
practically coincide.
In the inset of Fig.~\ref{Fig10} we compare the two versions of the RGM
(minimal and extended) as well as the HTE series  for $S=1/2$.
Obviously, both versions of the RGM agree well with each other. Note,
however, that this statement does not hold for larger values of $S$, cf. the
discussion in Sec.~\ref{sec:GS}. The
HTE approach for correlation
functions is also in good agreement with the RGM data down to $T \sim 0.4$.  

\begin{figure}[H]
\centering \includegraphics[scale=0.33,angle=-90]{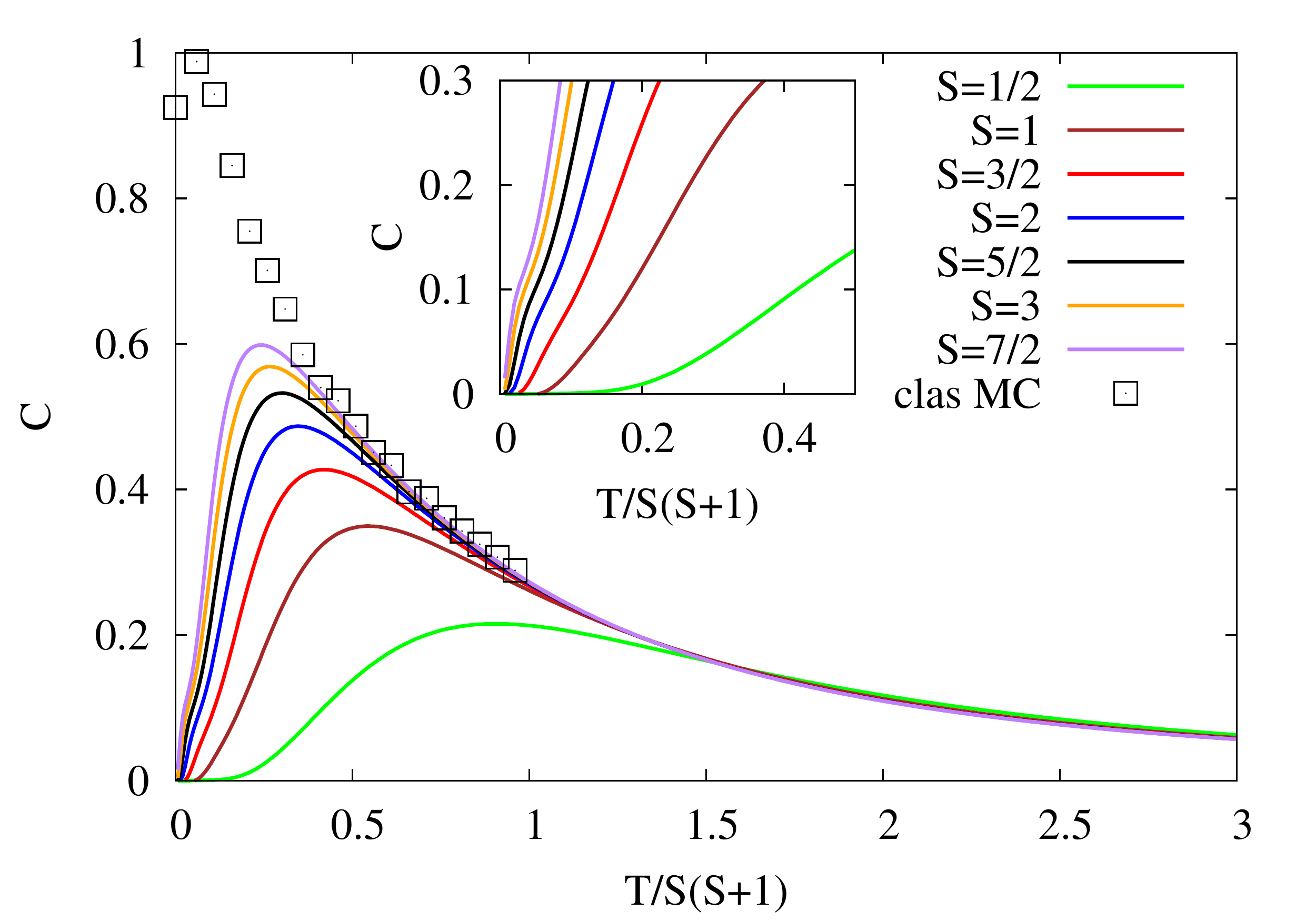} \protect\caption{
Main: Specific heat $C$ for various values of the spin quantum number $S$ as a function of the normalized temperature $T/S(S+1)$.
The Monte-Carlo data for the classical limit are taken from
\cite{Reimers1993,Huber2001}.
Inset:  Low-temperature behavior of $C$ using an enlarged scale.
}
\label{Fig11} 
\end{figure}

Now we turn to the specific heat.
For the extreme quantum case $S=1/2$ various methods 
provide indications for an additional low-temperature peak at about $T=0.1$
\cite{elstner1994,Nakamura1995,tomczak1996thermodynamical,Lhuillier_thermo_PRL2000,Bernu2005,Rigol2007,Munehisha2014,Shimokawa2016}
due to a set of low-lying singlet
states.
However, instead of a true maximum a shoulder-like hump may characterize the
low-$T$ profile of $C(T)$ \cite{Bernu2005,Xi-Chen2017}.
It is an open question whether for $S>1/2$ such a feature is still present. 
Our RGM approach does not show any unconventional feature in the temperature
profile of the specific heat at low $T$ for $S=1/2$ and $S=1$, cf.
Fig.~\ref{Fig11}.
For $S>1$ a weakly pronounced shoulder-like hump emerges (see the inset of
Fig.~\ref{Fig11}).      
We argue, that our RGM approach is not able to detect the subtle role of
low-lying 
excitations relevant for the low-temperature physics of the KHAF in the extreme
quantum limit of small spin $S$.
On the other hand,  in the limit of large $S$ the RGM data       
seem to approach the classical Monte-Carlo data \cite{Chalker1992,Huber2001} reasonably well.

\begin{figure}[H]
\centering \includegraphics[scale=0.33,angle=-90]{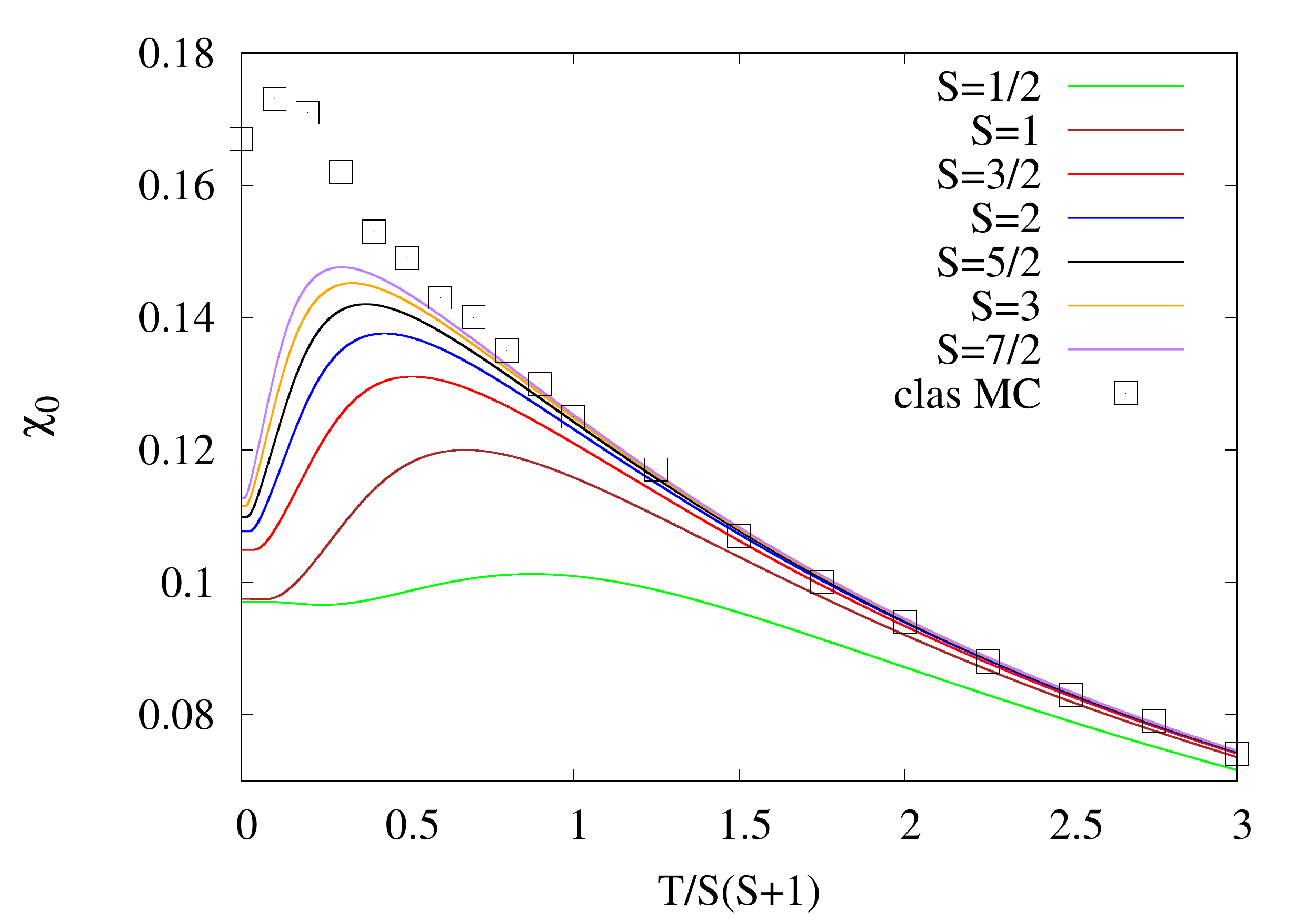} \protect\caption{
Uniform static susceptibility $\chi_{0}$ for various values of the spin $S$ as a function of the normalized temperature $T/S(S+1)$.
The Monte-Carlo data for the classical limit are taken from
\cite{Huber2001}.
}
\label{Fig12} 
\end{figure}

The temperature dependence of the static uniform susceptibility $\chi_0$ for
spin quantum numbers $S=1/2,1,\ldots,7/2$ is shown in Fig.~\ref{Fig12}.
Similar as for the specific heat there is a well-pronounced tendency to
shift the typical maximum in $\chi_0(T)$ towards lower values of $T/S(S+1)$
and to enlarge the height of the maximum as increasing $S$.
Again, in the limit of large $S$ the RGM data       
seem to approach the classical Monte-Carlo data 
\cite{Reimers1993,Huber2001} reasonably well.
The fact that $\chi_0(T=0)$ is finite, cf. also Fig.~\ref{Fig3},
 is in favor of a vanishing gap to magnetic excitations. There is an ongoing
controversial discussion of the  gap issue for the $S=1/2$ KHAF
\cite{Yan2011,Depenbrock2012,Laeuchli2011,Laeuchli2016,Iqbal2011,Iqbal2013,Xie2017}. 
However, we do not
claim, that our approach is accurate enough at low temperatures in the
quantum limit of small $S$ to provide   
reliable statements on the very existence of an excitation gap.

\subsubsection{Structure factor and correlation lengths}
\label{subsec_strufa}

\begin{figure}[H]
\hspace{-0.7cm}\includegraphics[scale=0.45]{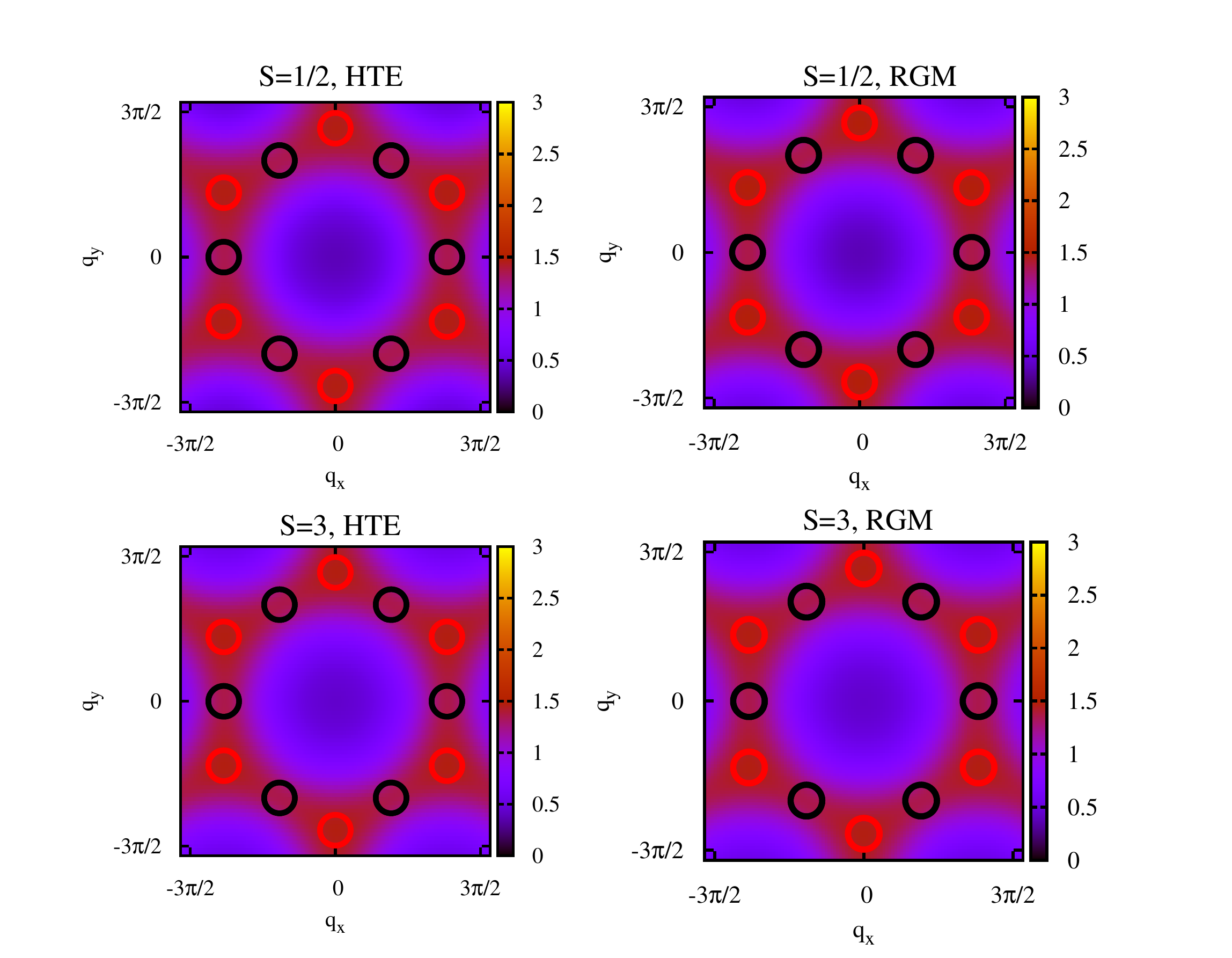} \protect\caption{
Intensity plot of the normalized structure factor ${\cal S}(\mathbf{q})/S(S+1)$ within  the
first and
extended Brillouin zones [cf. Fig.~\ref{Fig8}(a)] for $S=1/2$ and $S=3$ at
$T/S(S+1)=1.3$ (left: 9th order HTE, right: RGM).  
The red (black) circles indicate the expected
maxima for a
classical $\sqrt{3}\times \sqrt{3}$ ($q=0$) state.
}
\label{Fig13} 
\end{figure}

To get more insight in the magnetic ordering of the KHAF at finite
temperatures we investigate the structure factor and the correlation lengths.
Some information on magnetic SRO has already been provided in
Fig.~\ref{Fig10}. 
First we show in Fig.~\ref{Fig13} an intensity  plot of the static 
structure factor ${\cal S}(\mathbf{q})/S(S+1)$ for 
$S=1/2$ and $S=3$ for  $T/S(S+1)=1.3$ and compare RGM and HTE.
The overall impression is that the RGM and HTE approaches yield very similar
intensity plots  of ${\cal S}(\mathbf{q})/S(S+1)$.  The
characteristic hexagonal bow-tie pattern (i.e., the intensity is concentrated
along the edge of the extended Brillouin zone), which was found at $T=0$,
cf. Fig.~\ref{Fig8}, is still present at
$T/S(S+1)=1.3$.

\begin{figure}[ht!]
\centering \includegraphics[scale=0.7]{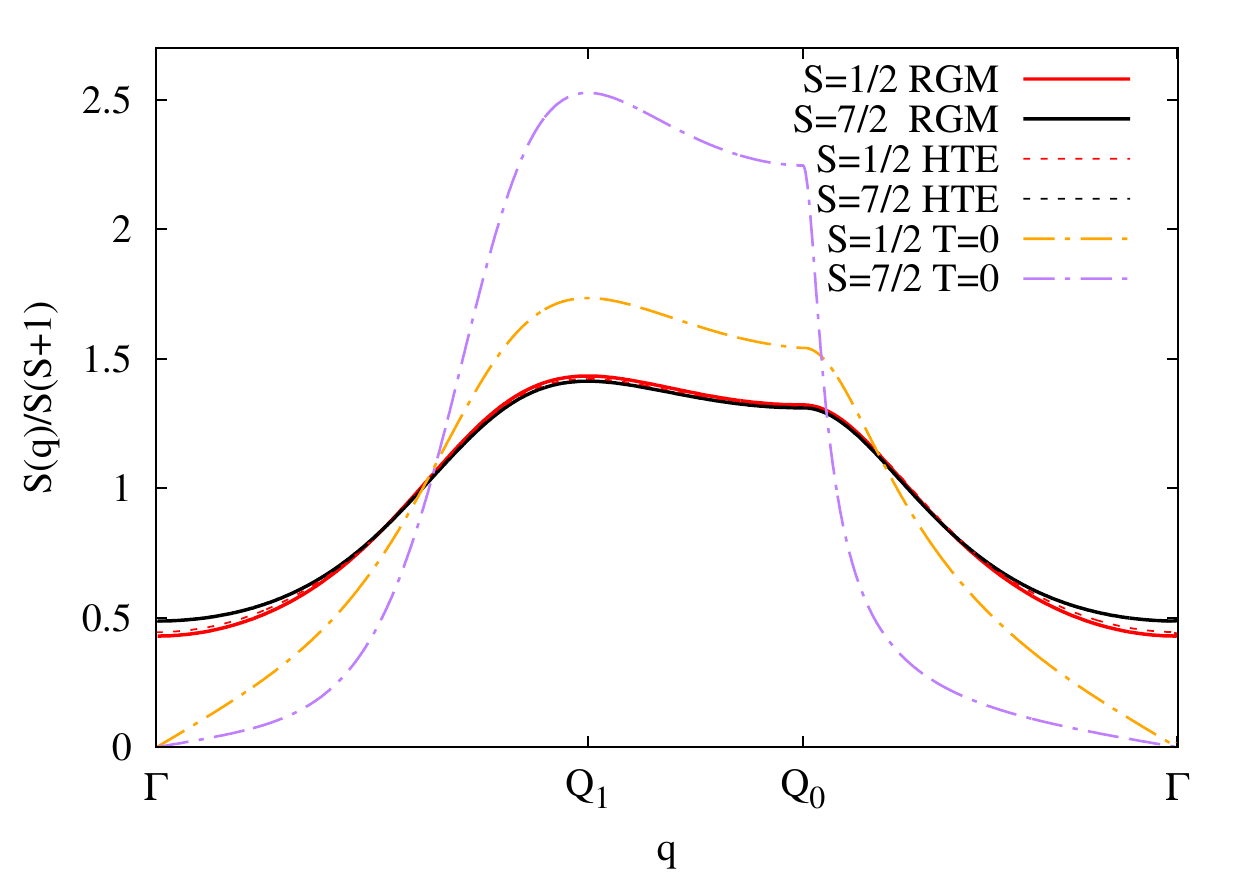} \protect\caption{
RGM and HTE data for the normalized structure factor ${\cal
S}(\mathbf{q})/S(S+1)$ along the path $\Gamma \to {\mathbf{Q}_1} \to {\mathbf{Q}_0}
\to \Gamma $ 
for  $S=1/2$ and $S=7/2$ at $T/S(S+1)=1.5$. For
comparison we also present RGM data for $T=0$ (dashed-dotted lines).
}
\label{Fig14} 
\end{figure}

Next we show in Fig.~\ref{Fig14} the static 
structure factor ${\cal S}(\mathbf{q})/S(S+1)$ along the path $\Gamma \to {\mathbf{Q}_1}
\to {\mathbf{Q}_0}
\to \Gamma $ for  $S=1/2$ and $S=7/2$ for  $T/S(S+1)=1.5$ (RGM and HTE)
and $T=0$ (only RGM, see also
Fig.~\ref{Fig9}).
Obviously, the temperature $T/S(S+1)=1.5$ is already large enough, such
that all four curves are very close to each other.  
Although, the weakening of magnetic ordering by thermal fluctuations is
evident,  the overall shape of the finite-temperature curves is similar to the GS
curves, especially  the maxima at ${\mathbf{Q}_1}$
($\sqrt{3}\times\sqrt{3}$ state)
and at ${\mathbf{Q}_0}$ ($q=0$ state) are still present, and
${\cal S}(\mathbf{Q}_1)>
{\cal S}(\mathbf{Q}_0)$.

\begin{figure}
\centering
\includegraphics[scale=0.35,angle=-90]{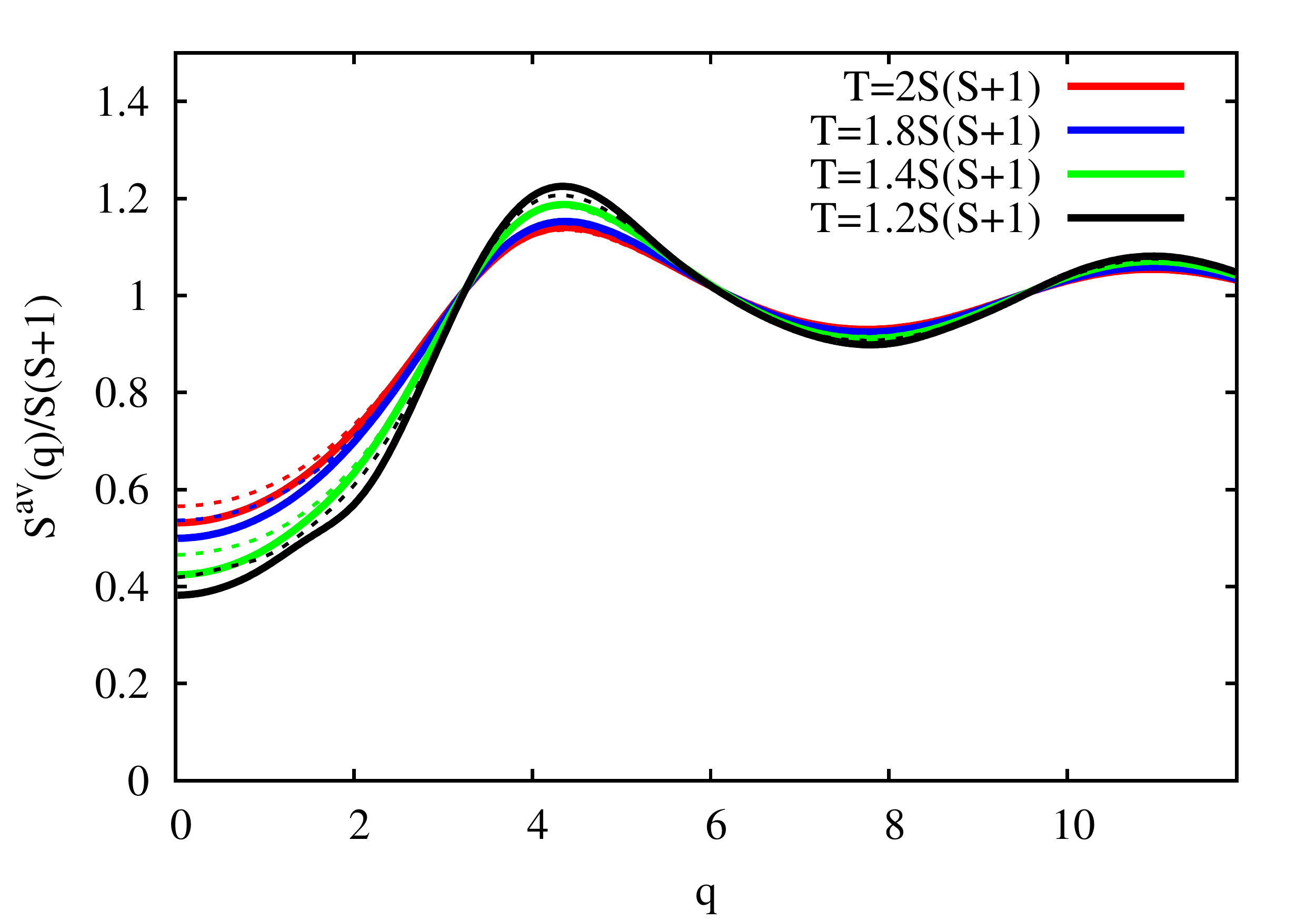} \protect\caption{
Normalized powder-averaged structure factor ${\cal S}^{\rm
av}(|\mathbf{q}|)/S(S+1)$ for spin $S=1/2$ (solid lines) and $S=7/2$ (dashed
lines) using HTE of 9th order. 
}
\label{Fig15} 
\end{figure}

In experiments, often neutron scattering on powder samples are performed,
see, e.g. \cite{herbertsmithite2009}.
Hence, we also present the powder-averaged structure factor ${\cal S}^{\rm
av}(|\mathbf{q}|)/S(S+1)$, i.e., we integrate over
all points at equal $q=|\mathbf{q}|$. 
We show  HTE data for ${\cal S}^{\rm av}(|\mathbf{q}|)/S(S+1)$ for $S=1/2$ and
$S=7/2$ at various temperatures in Fig.~\ref{Fig15}. 
The first broad maximum at $|\mathbf{q}| \sim 4.36$  corresponds
to short-ranged antiferromagnetic correlations and its position
is in good agreement with experiments on Herbertsmithite
\cite{herbertsmithite2009}. (Note that the separation of NN copper ions in Herbertsmithite
is $a=3.4$\AA, here we use $a=1$.) 
While the influence of $T$ on the height of the maxima in  $S^{\rm
av}(|\mathbf{q}|)$ is recognizable, the position of the maxima is almost
independent of $T$. 
Thus, from Fig.~\ref{Fig15} and Fig.~\ref{Fig14} one can conclude
that the type of  magnetic SRO found at pretty high  temperatures $T/S(S+1)
> 1$ indicate a possible magnetic ordering at low temperatures.      

\begin{figure}[H]
\centering \includegraphics[scale=0.33,angle=-90]{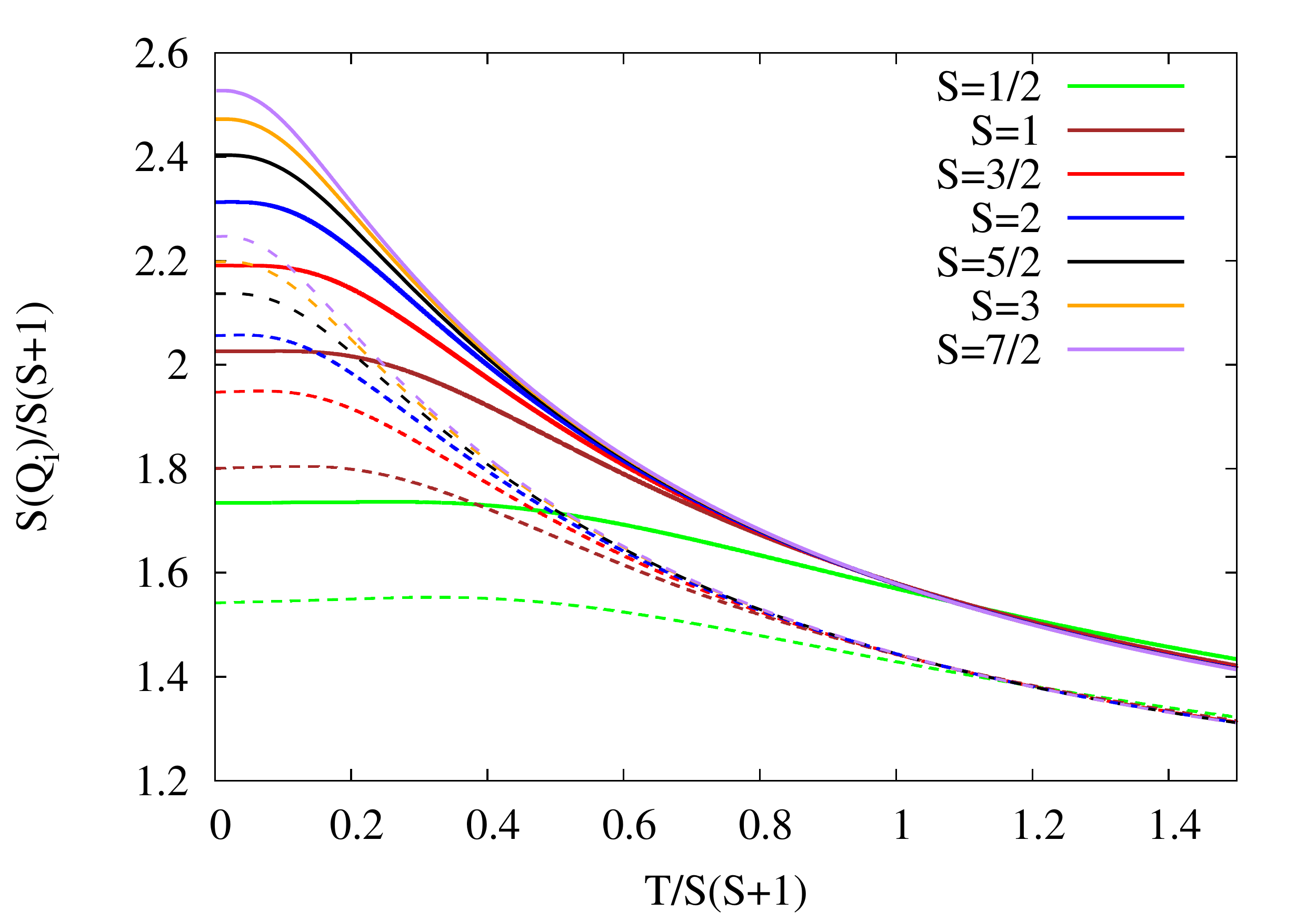} \protect\caption{
RGM data for the normalized structure factor ${\cal S}(\mathbf{Q}_i)/S(S+1)$
at the magnetic wave vector $\mathbf{Q}_i$ (dashed -
$\mathbf{Q}_0$; solid - $\mathbf{Q}_1$) for various values of the spin
$S$
as a function of the normalized temperature $T/S(S+1)$.
}
\label{Fig16} 
\end{figure}

\begin{figure}[H]
\centering \includegraphics[scale=0.33,angle=0]{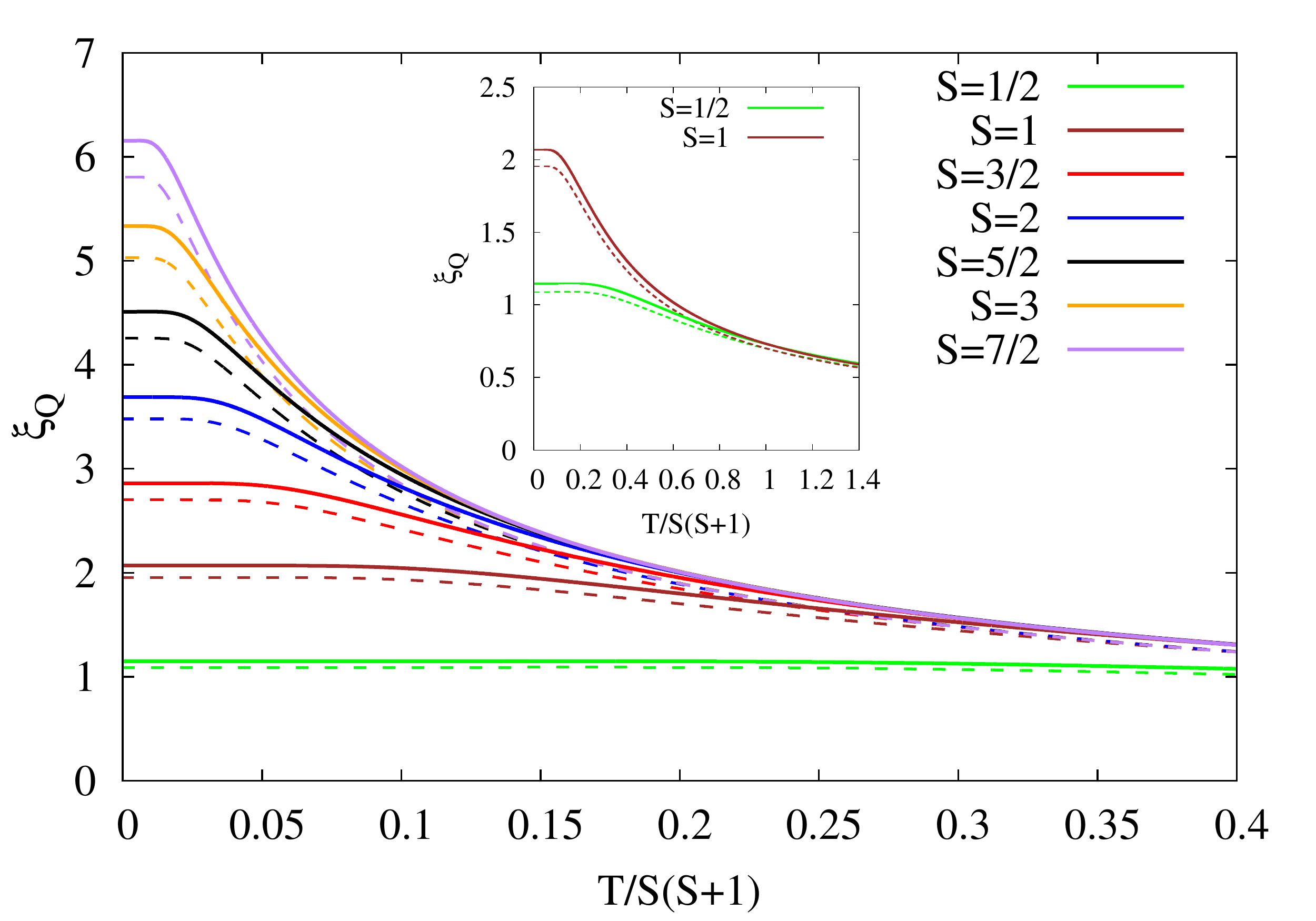} \protect\caption{
Main: RGM data for the correlation length $\xi_{\mathbf{Q}_i}$ (dashed -
$\mathbf{Q}_0$; solid - $\mathbf{Q}_1$) for various values of the spin
$S$ as a function of the normalized temperature $T/S(S+1)$. Inset:
Correlation lengths
$\xi_{\mathbf{Q}_i}$ (dashed - $\mathbf{Q}_0$; solid - $\mathbf{Q}_1$) for $S=1/2$ and $S=1$
using an enlarged y-axis.}
\label{Fig17} 
\end{figure}

\begin{figure}[H]
\centering \includegraphics[scale=0.33,angle=-90]{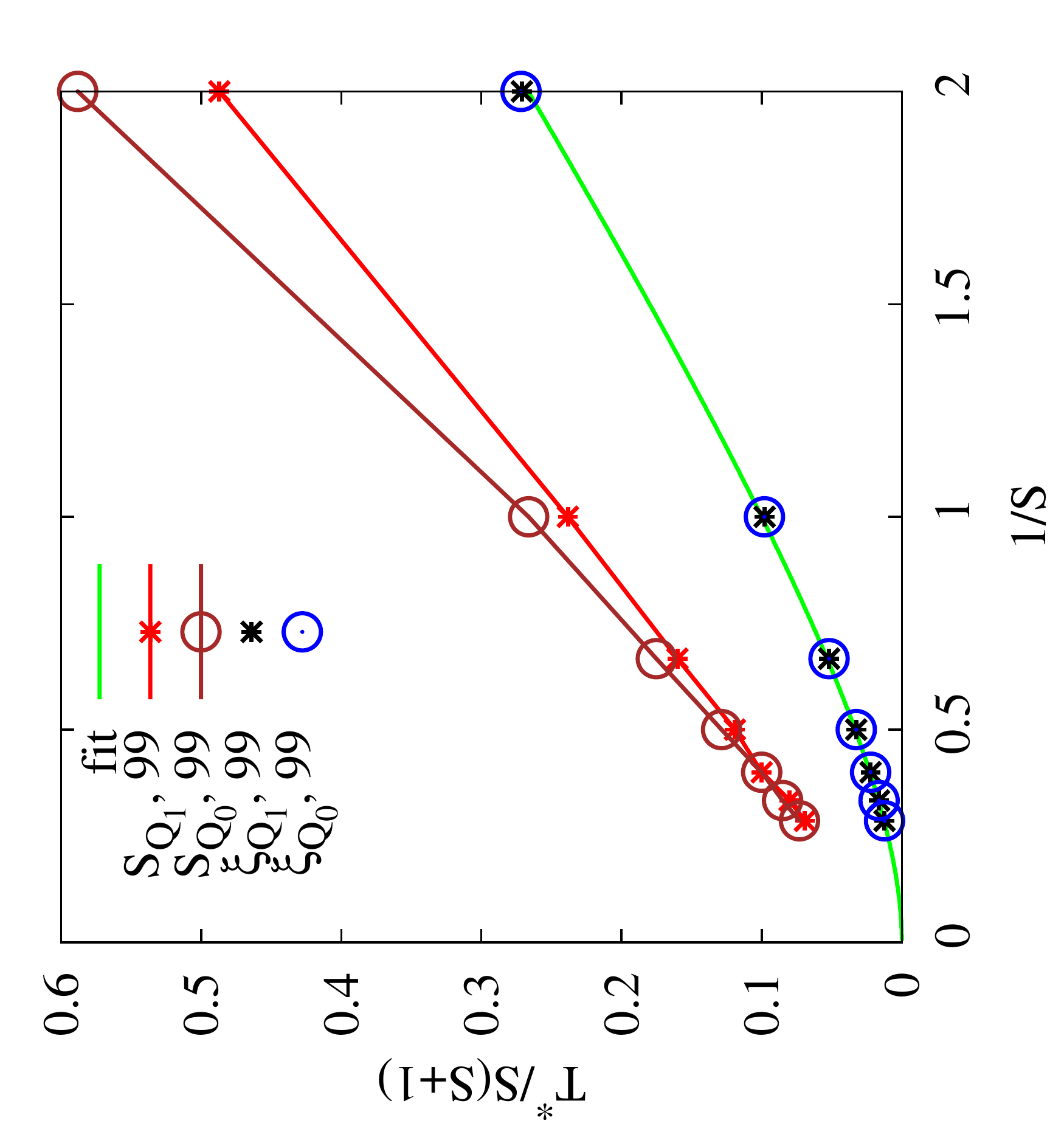} \protect\caption{
RGM data for the temperature region $T^*/S(S+1)$ of almost almost constant
($p=99\%$, see text)
correlation lengths and structure factors
 as a function of  $1/S$. The green line corresponds the fit function 
$f(S)=0.2/S(S+1)$.
}
\label{Fig18} 
\end{figure}

Last but not least we discuss the temperature dependence of the structure
factors at the magnetic wave
vectors $\mathbf{Q}_0$ ($q=0$ state)  and $\mathbf{Q}_1$ ($\sqrt{3}\times \sqrt{3}$
state) and of the corresponding
correlation lengths $\xi_{\mathbf{Q}_0}$
and $\xi_{\mathbf{Q}_1}$, see Figs.~\ref{Fig16} and \ref{Fig17}. 
First we note  that the  $\sqrt{3}\times \sqrt{3}$ SRO  is more pronounced 
than the $q=0$ SRO
for
all temperatures $T \ge 0$, i.e.,  ${\cal S}(\mathbf{Q}_1)|_{T} >{\cal S}(\mathbf{Q}_0)|_{T}$ and 
$\xi_{\mathbf{Q}_1}(T) >\xi_{\mathbf{Q}_0}(T)$ (cf. also Figs.~\ref{Fig7} and \ref{Fig9}
for the GS).   
As increasing $S$ the SRO becomes  more distinct. Only at temperatures
$T/S(S+1) \gtrsim 1$ the curves for different $S$ collapse to one universal
curve, cf. \cite{Lohmann2014}.  
At low temperatures $T < T^*$  we find a plateau-like behavior in the
correlation lengths and the structure factors,
$\xi_{\mathbf{Q}_i}|_{T<T^*} \approx \xi_{\mathbf{Q}_i}|_{T=0}$ and 
${\cal S}(\mathbf{Q}_i)|_{T<T^*} \approx {\cal S}(\mathbf{Q}_i)|_{T=0}$.
The region of almost constant correlation lengths and structure factors
is largest for $S=1/2$  and it shrinks
noticeably as
increasing $S$ approaching zero in the classical limit ($\lim_{S \to \infty} T^*/S(S+1) =
0$).
To define a reasonable estimate of $T^*$ we chose that value of $T$, where
correlation lengths and the structure factors reach $p=99\%$ of its GS values.
The corresponding data are shown in Fig.~\ref{Fig18}.
We mention that for the correlation lengths
the relation $T^*=a/S(S+1)$ describes the plotted behavior accurately, where
$a=0.2$ for $p=99\%$. (Note that the prefactor $a$ increases only slightly to
$a=0.28$ as
changing $p$ to $p=95\%$.)  
We may argue that  below $T^*$ the quantum fluctuations are more important
than
thermal fluctuations.

\section{Summary}
\label{sec:sum} 
We use two methods to discuss the thermodynamic properties of the
kagome Heisenberg antiferromagnet with arbitrary spin $S$, namely the
rotational invariant Green's function method (RGM) and the high-temperature
expansion (HTE).
Within the RGM we consider GS as well as finite-temperature properties,
whereas the HTE is restricted to $T/S(S+1) \gtrsim 1$.  
Within the RGM approach the model does not exhibit magnetic LRO for all
values of $S$.
In the extreme quantum case $S=1/2$ the
zero-temperature  correlation
length $\xi(T=0)$ is only of the order of the nearest-neighbor separation.
As increasing  $S$  the correlation
length $\xi(T=0)$ grows 
according to a power-law in $1/S$.  
We found that the so-called
$\sqrt{3}\times\sqrt{3}$ SRO is favored versus the
$q=0$ SRO for all values of $S$. 
It is worth mentioning that other methods specifically designed for the GS
\cite{Goetze2011,Goetze2015,Oitmaa2016,Liu2016}
indicate that GS LRO may appear for $S\ge 3/2$.
As known from previous studies 
the rotational invariant decoupling in the RGM scheme may overestimate the
tendency to suppress magnetic order, cf.
\cite{Barabanov94,Ihle2001,Schmalfus2004,Haertel2013,Mikheyenkov2013} and
references
therein.

As typical for two-dimensional Heisenberg antiferromagnets, the specific heat and the uniform 
susceptibility exhibit a maximum related to the size of the exchange coupling
$J$.
For both quantities, with growing $S$ this maximum moves towards lower values of $T/S(S+1)$ 
and its height increases.
In the limit of large $S$ the RGM data       
approach the classical curves.

The structure factor ${\cal S}(\mathbf{q})$ shows
two maxima at magnetic wave vectors $\mathbf{q}={\mathbf{Q}_i}, i=0,1$,
corresponding to the $q=0$ and $\sqrt{3}\times\sqrt{3}$ state, where 
 ${\cal
S}(\mathbf{Q}_1)>{\cal S}(\mathbf{Q}_0)$ holds for all values of $S$ and
all temperatures $T \ge 0$.
In a finite low-temperature region $T < T^*\approx a/S(S+1), a \approx 0.2$, the magnetic SRO   
is quite stable against thermal fluctuations, i.e., the correlation lengths and the structure factors
${\cal
S}(\mathbf{Q}_1)$ and ${\cal S}(\mathbf{Q}_0)$
are almost independent of $T$.
The powder-averaged structure factor  ${\cal S}(|\mathbf{q}|)$ 
exhibits a 
 broad maximum related
to short-ranged antiferromagnetic correlations and its position
is in good agreement with experiments on powder samples of Herbertsmithite
\cite{herbertsmithite2009}.

\section*{Acknowledgments}

The authors thank D. Ihle and Paul McClarty for valuable hints.

\bibliography{JR_RGM}

\end{document}